%% file: manuscript.tex
\documentclass[journal]{IEEEtran}
\usepackage{stmaryrd}
\usepackage{amsmath}

\usepackage{amsmath}
\usepackage{amsfonts}%
\usepackage{amssymb}
\usepackage{mathtools}
\usepackage{algorithmicx}
\usepackage{algpseudocode}
\usepackage{algorithm}
\usepackage{physics}
\usepackage{relsize}
\usepackage{gensymb}
\algdef{SE}[DOWHILE]{Do}{doWhile}{\algorithmicdo}[1]{\algorithmicwhile\ #1}%

\usepackage{flushend}

\usepackage{array}
\usepackage{subcaption}
\usepackage[font=normalsize,labelformat=simple]{subfig}

\usepackage{blindtext}
\usepackage{tikz}
\usepackage{amsthm}

\newtheorem{definition}{Definition}

\usepackage{titlesec}

\titlespacing\section{5pt}{12pt plus 4pt minus 2pt}{4pt plus 2pt minus 2pt}
\titlespacing\subsection{0pt}{12pt plus 4pt minus 2pt}{3pt plus 2pt minus 2pt}

\begin{document}

\title{Progressive-Proximity Bit-Flipping for Decoding Surface Codes}

\author{Michele Pacenti,~\IEEEmembership{Graduate Student Member,~IEEE,}\ Mark F. Flanagan,~\IEEEmembership{Senior Member,~IEEE,}
Dimitris Chytas,
and Bane Vasi\'c,~\IEEEmembership{Fellow,~IEEE}

\thanks{This work has been conducted as a part of the CoQREATE program which is funded by the National Science Foundation under grant ERC-1941583 and Science Foundation Ireland under the US-Ireland R\&D Partnership Programme under grant SFI/21/US-C2C/3750.
The work of Michele Pacenti, Dimitris Chytas and Bane Vasi\'c is also supported by the NSF under grants CIF-1855879, CIF-2106189, CCF-2100013, ECCS/CCSS-2027844, ECCS/CCSS-2052751, and in part by Jet Propulsion Laboratory, California Institute of Technology, under a contract with the National Aeronautics and Space Administration and
funded through JPL’s Strategic University Research Partnerships
(SURP) program. Bane Vasi\'{c} has disclosed an outside
interest in his startup company Codelucida to The University
of Arizona. Conflicts of interest resulting from this interest are
being managed by The University of Arizona in accordance
with its policies.}

\thanks{Michele Pacenti, Dimitris Chytas and Bane Vasi\'{c} are with the Department of Electrical and Computer Engineering, The University of Arizona, Tucson, AZ 85721, USA  (e-mail:  mpacenti@arizona.edu; dchytas@arizona.edu; \mbox vasic@ece.arizona.edu). }

\thanks{Mark F. Flanagan is with the School of Electrical and Electronic Engineering, University College Dublin, Belfield, Dublin 4, Ireland (e-mail: mark.flanagan@ieee.org). }
}

\maketitle

\begin{abstract}
Topological quantum codes, such as toric and surface codes, are excellent candidates for hardware implementation due to their robustness against errors and their local interactions between qubits. However, decoding these codes efficiently remains a challenge: existing decoders often fall short of meeting requirements such as having low computational complexity (ideally linear in the code's blocklength), low decoding latency, and low power consumption. In this paper we propose a novel bit-flipping (BF) decoder tailored for toric and surface codes. We introduce the proximity vector as a heuristic metric for flipping bits, and we develop a new subroutine for correcting degenerate multiple errors on adjacent qubits. Our algorithm has quadratic complexity growth and it can be efficiently implemented as it does not require operations on dynamic memories, as do state-of-art decoding algorithms such as minimum weight perfect matching or union find. The proposed decoder shows a decoding threshold of $7.5\%$ for the 2D toric code and $7\%$ for the rotated planar code over the binary symmetric channel.
\end{abstract}

\begin{IEEEkeywords}
Surface codes, topological codes, bit flipping, decoding algorithm, quantum error correction.
\end{IEEEkeywords}

\IEEEpeerreviewmaketitle

\section{Introduction}
%
%
%
%
\IEEEPARstart{Q}{uantum} computers make use of the principles of quantum mechanics to perform computations. Quantum states are fragile and very sensitive to errors, and thus it is crucial to implement quantum error correction techniques to protect quantum information. A very important class of quantum codes are topological codes, specifically surface and toric codes \cite{dennis_topological_2002}, as they can be implemented on a planar quantum chip. Decoding of surface codes is typically performed using the minimum-weight-perfect-matching (MWPM) decoder \cite{fowler_minimum_2014}; although MWPM provides excellent decoding performance, its computational complexity makes it infeasible for large-scale implementation. Because of this, many alternative approaches have been studied to lower the decoding complexity of surface codes. The most promising alternative is the union-find (UF) decoder \cite{delfosse_almost-linear_2021}, although many other classes of decoders have been proposed, such as tensor network decoders \cite{bravyi_efficient_2014} (which are implementations of maximum likelihood decoding), re-normalization group decoders \cite{duclos-cianci_fast_2010}, neural network based decoders \cite{Gicev_2023}, maxSAT decoders \cite{berent_decoding_2023} and cellular-automaton decoders \cite{herold_cellular-automaton_2015}. On the other hand, message passing decoders such as belief propagation (BP) have shown to be effective for decoding surface codes when paired with MWPM and UF \cite{higgott_improved_2023}, or with post-processing techniques such as ordered statistics decoding (BP-OSD) \cite{panteleev_degenerate_2021}. Also, serial scheduling and re-initialization of BP has been shown to provide good decoding performance \cite{kuo_exploiting_2022}. 

Bit flipping (BF) decoders are a class of iterative decoding algorithms known to be very fast and efficient \cite{ivanis_suspicion_2022}, although generally they provide lower performance than BP. In general, BF decoders do not perform well on surface codes, mainly because of the very low column weight of the parity check matrix, and also because of \textit{error degeneracy}, which refers to the property of quantum codes where multiple error patterns of the same weight can correspond to the same syndrome. Nevertheless, because of its extremely low complexity, BF is still attractive in the scenario of decoding topological codes; indeed, the latency constraint that a quantum decoder should meet is very tight, to prevent the so-called \textit{backlog} problem \cite{holmes_nisq_2020}. Moreover, it is essential that the decoder has a low power consumption, as it has to be embedded in a cryogenic environment with a strict power budget \cite{krinner_engineering_2019}. This makes hardware implementation of the decoder a crucial aspect; unfortunately, state-of-art decoders such as MWPM or UF do not allow an efficient hardware implementation, mainly because they make use of complex data structures, which require random memory access and dynamic memory allocation, which are known to be less efficient as they introduce latency and increase circuit complexity as well as power consumption. In contrast, the BF algorithm, due to its simplicity, only requires static memories with fixed access, thus having a distinct advantage for hardware implementation.

In this paper, we develop a BF algorithm which is capable of decoding surface codes. In the proposed approach, instead of considering the number of unsatisfied checks as in conventional BF, each bit is assigned a heuristic weight which is the entry of what we call \textit{proximity vector}. The proximity vector is the sum of different contributions called \textit{individual influences}, and each individual influence is associated with an unsatisfied check. Ultimately, bits connected with checks with low entries in the proximity vector will be flipped first, while bits connected with checks with high entries in the proximity vector will be flipped last. After each flip, the proximity vector is updated in an efficient manner. To deal with high-weight consecutive errors, \textit{\textit{i.e.}}, errors occurring on a set of adjacent qubits, which are particularly harmful for iterative decoders, we design an iterative matching procedure that runs after BF and that is able to correct these errors. We show that our decoder has an asymptotic complexity of $\mathcal{O}(n^2)$, where $n$ is the code's blocklength, and we provide simulation results that show a comparison, in terms of performance, with MWPM, UF and traditional BF. Although our decoder is not able to correct up to the minimum distance of the code, we show that it has a good performance while also being more feasible for hardware implementation than UF and MWPM.

The rest of the paper is organized as follows. In Section \ref{sec:preliminaries} we introduce the preliminaries of quantum error correction. In Section \ref{sec:state_of_art} we present an overview of the main decoding algorithms for topological codes. In Section \ref{sec:proximity} we define the proximity vector and how it can be computed efficiently.  Section \ref{sec:decoder} provides a detailed description of our proposed decoder. In Section \ref{sec:complexity} we carry out a complexity analysis and a comparison with other state-of-art decoders. In Section \ref{sec:hardware} we analyze the hardware implementation aspects for UF, MWPM and the proposed decoder. Finally, Section \ref{sec:results} presents simulation results.


\section{Preliminaries}
\label{sec:preliminaries}

\subsection{Stabilizer formalism}
Consider the $n$-fold Pauli group,
\[
\mathcal{G}_n  \triangleq \{cB_1 \otimes \cdots \otimes cB_n: c \in \{\pm 1, \pm i \}, B_j \in \{I, X, Y, Z\} \},
\]
where $I=\begin{bsmallmatrix}
    1 & 0\\
    0 & 1
\end{bsmallmatrix},X=\begin{bsmallmatrix}
    0 & 1\\
    1 & 0
\end{bsmallmatrix},Y=\begin{bsmallmatrix}
    0 & -i\\
    i & 0
\end{bsmallmatrix},Z=\begin{bsmallmatrix}
    1 & 0\\
    0 & -1
\end{bsmallmatrix}$. Every non-identity Pauli operator $P \in \mathcal{G}_n$ has eigenvalues $\pm 1$ and any two Pauli operators in $\mathcal{G}_n$ either commute or anti-commute with each other.   The weight of a Pauli operator $P$ is defined as the number of non-identity elements in the tensor product.

By dropping the phase factor $c$, the Pauli group $\mathcal{G}_n$ is isomorphic to $\mathbb{F}_2^{2n}$ \cite{panteleev_degenerate_2021}, such that
\begin{equation}
\label{eq:bin_map}
    c\underset{i=1}{\overset{n}{\mathlarger{\mathlarger{\boldsymbol{\otimes}}}}}X^{x_i}Z^{z_i} \mapsto (x_1,...,x_n\ |\ z_1,...,z_n).
\end{equation}
In this representation, the commutation relation between two Pauli operators $\mathbf{p}_1 = (\mathbf{x_1}|\mathbf{z_1})$ and $\mathbf{p}_2=(\mathbf{x_2}|\mathbf{z_2})$ can be computed by checking that the \textit{symplectic inner product} is zero:
\begin{equation}
\label{eq:symplectic}
    \mathbf{x_1}\mathbf{z_2}^T + \mathbf{z_1}\mathbf{x_2}^T=0\ \mod 2.
\end{equation}

A stabilizer group $\mathcal{S}$ is an Abelian subgroup of $\mathcal{G}_n$, and
an $\llbracket n,k,d \rrbracket $ stabilizer code  is a $2^k$-dimensional subspace $\mathcal{C}$ of the Hilbert space $(\mathbb{C}^2)^{\otimes n}$ that satisfies the condition $S_i\ket{\Psi} = \ket{\Psi},\ \forall\ S_i\in \mathcal{S}, \ket{\Psi}\in \mathcal{C}$. Thus, the code $\mathcal{C}$ is defined as the common $+1$-eigenspace of the stabilizer group $\mathcal{S}$. The stabilizer group $\mathcal{S}$ is generated by a set of $n-k$ independent generators $ S_1,...,S_{n-k}$ that can be represented using a matrix $\mathbf{S}$, called the \emph{stabilizer matrix}, whose $(i,j)$ element is given by the Pauli operator corresponding to the $j$-th qubit in the $i$-th stabilizer. The minimum distance $d$ is defined as the minimum weight of an element of $ N(\mathcal{S}) \setminus \mathcal{S}$, where $N(\mathcal{S})$ is the normalizer\footnote{The normalizer subgroup $N(\mathcal{S})$ of a group $\mathcal{S}$ is the subgroup of $\mathcal{S}$ which is invariant under conjugation.} of $\mathcal{S}$. Elements in $N(\mathcal{S}) \setminus \mathcal{S}$ are also called \textit{logical operators}.

Applying the mapping (\ref{eq:bin_map}) to the stabilizer matrix $\mathbf{S}$, we obtain an $(n-k) \times 2n$ binary matrix:
\begin{equation}
    \mathbf{H} = [\mathbf{H}_X\ |\ \mathbf{H}_Z]
\end{equation}
which we call the \textit{parity check matrix} of $\mathcal{C}$. Since the corresponding stabilizers of $\mathbf{S}$ commute with each other, it is easy to check using (\ref{eq:symplectic}) that
\begin{equation}
\label{eq:commutativity}
    \mathbf{H}_X\mathbf{H}_Z^T + \mathbf{H}_Z\mathbf{H}_X^T = \mathbf{0}\ \mod 2.
\end{equation}

\subsection{Error model and syndrome}
\label{sec:noise_model}
In the depolarizing error model, errors are Pauli operators such that $\mathbf{e}\in \{I,X,Z,Y\}^{\otimes n}$; each error on the $i$-th qubit $e_i$ can either be a bit flip ($X$), a phase flip ($Z$) or both ($Y$), each with probability $\epsilon/3$, while the probability of no error ($I$) is equal to $1-\epsilon$. Using the Pauli-to-binary mapping of (\ref{eq:bin_map}), a Pauli error $\mathbf{e}$ can be also be modeled as a $1 \times 2n$ binary vector $\mathbf{e} = [\mathbf{e}_X\ \mathbf{e}_Z]$. Given a Pauli error $\mathbf{e}$, the corresponding syndrome $\mathbf{s} \in \{0,1\}^{n-k}$ can be computed using the symplectic inner product such that
\begin{equation}
    \label{eq:syndrome}
    \mathbf{s} = \mathbf{e}_X\mathbf{H}_Z^T + \mathbf{e}_Z\mathbf{H}_X^T\ \mod 2.
\end{equation}
Obviously, if $\mathbf{e}\in \mathcal{S}$ we have $\mathbf{s}=\mathbf{0}$. More generally, any Pauli error can be expressed as a combination of a \textit{true error} $\mathbf{e}_t$, a stabilizer $\mathbf{S}_i$ and a logical operator $\mathbf{L}_j$ such that $\mathbf{e} = \mathbf{e}_t + \mathbf{S}_i + \mathbf{L}_j$, with $\mathbf{S}_i \in \mathcal{S}$ and $\mathbf{L}_j \in N(\mathcal{S}) \setminus \mathcal{S}$; since, by definition, elements of $\mathbf{L}_j$ and $\mathcal{S}_i$ commute with each other (thus, their syndrome is zero), the syndrome $\mathbf{s}$ is only dependent on the true error $\mathbf{e}_t$; moreover, any error estimate of the type $\hat{\mathbf{e}} = \mathbf{e}_t + \mathbf{S}_i$ is a valid error estimate. This phenomenon is known as \textit{error degeneracy}. It is well known that, if the minimum distance of a stabilizer code is much higher than the weight of its stabilizer elements, there will be many degenerate errors of the same weight \cite{kovalev_fault_2013}.



\subsection{Calderbank-Shor-Steane codes}
An $\llbracket n, k_X-k_Z, d \rrbracket $ Calderbank-Shor-Steane (CSS) code $\mathcal{C}$ is a stabilizer code constructed using two classical  $[n,k_X,d_X]$ and $[n,k_Z,d_Z]$ codes $C_X$ and $C_Z$, respectively, where $d\geq~\mathrm{min}\{d_X,d_Z\}$ and $C_Z \subset C_X$ \cite{calderbank_good_1996}. Note that $k_X$, $k_Z$  and $d_X$, $d_Z$ correspond to the dimensions and minimum distances of $C_Z$ and $C_X$, respectively. The parity check matrix of the CSS code $\mathcal{C}$ has the form 
\begin{equation}
    \mathbf{H}=\begin{bmatrix}
    \mathbf{H}_Z & \mathbf{0}\\
    \mathbf{0} & \mathbf{H}_X \end{bmatrix},
\end{equation}
where $\mathbf{H}_X$ and $\mathbf{H}_Z$ are the parity check matrices of $C_X$ and $C_Z$, respectively, and the commutativity condition of (\ref{eq:commutativity}) reduces to $\mathbf{H}_X\mathbf{H}_Z^T = \mathbf{0} \mod 2$. To correct depolarizing errors on the qubits, a syndrome $\mathbf{s}$ is computed such that 
\begin{equation}
    \mathbf{s} = [\mathbf{s}_X\ \mathbf{s}_Z],
\end{equation}
where $\mathbf{s}_X = \mathbf{e}_X \mathbf{H}_Z^T\mod 2$ and $\mathbf{s}_Z = \mathbf{e}_Z \mathbf{H}_X^T\mod 2$. Because of the structure of $\mathbf{H}$, $X$ and $Z$ errors can be corrected independently using $\mathbf{H}_Z$ and $\mathbf{H}_X$, respectively. 
In this paper we consider a bit-flip channel, where each qubit experiences an $X$ error with probability $p$, and remains correct with probability $1-p$. In other words, we fix $\mathbf{e}_Z=\mathbf{0}$, while elements in $\mathbf{e}_X$ can be 1 with probability $p$ or 0 with probability $1-p$. Hence we only  consider $\mathbf{H}_Z$ and $\mathbf{s}_X$, which for simplicity we will denote as $\mathbf{H}$ and $\mathbf{s}$, respectively. Note that the bit-flip channel that we consider and the depolarizing error model are closely related: indeed, assuming that $X$ and $Z$ errors are decoded separately, we can model the depolarizing channel as two binary symmetric channels with probability of error $p_x=p_z=\frac{2}{3}\epsilon$; assuming that $d_X=d_Z$, is possible to switch from the logical error rate curve over a binary symmetric channel to the logical error rate under depolarizing noise by re-scaling it of a factor of $3/2$ \cite{mackay_sparse-graph_2004}.

It is convenient to introduce the notion of \textit{Tanner graph} \cite{tanner_recursive_1981}. A Tanner graph is a bipartite graph defined from $\mathbf{H}$, such that it has two sets of nodes $V = \{v_1,v_2,...,v_n\}$ and $C = \{c_1,c_2,...,c_{n-k}\}$ called \textit{variable nodes} and \textit{check nodes}, respectively, and there is an edge between $v_j$ and $c_i$ if and only if $h_{i,j}=1$. A check node $c_i$ is said to be \textit{satisfied} if $s_i=0$, and \textit{unsatisfied} if $s_i=1$. The degree of a variable or check node is defined as the number of its incident edges. We define the distance between two nodes $i$ and $j$ to be the number of variable nodes belonging to the shortest path between $i$ and $j$.

\subsection{Surface and toric codes}
\begin{figure}[!htb]
    \centering
    \input{tikz_figures/lattice_numbering.tikz}
    \caption{Tanner graph for the $\llbracket 9,1,3 \rrbracket $ rotated planar code.  Circles (with black labels) correspond to variable nodes, while squares (with red labels) correspond to check nodes.}
    \label{fig:lattice_numbering}
\end{figure}
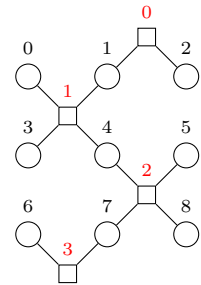

Surface codes \cite{dennis_topological_2002} are a widely known class of CSS codes. These are derived from the tessellation of the topological surface in squares, in such a way as to form a lattice. Generally, a surface code is characterized by an $L \times L$ lattice, where $L$ is the size of the horizontal (or vertical) dimension. In the lattice we can identify vertices as $X$ checks, edges as qubits and squares as $Z$ checks. Although there are several types of surface codes, in this paper we focus on the rotated planar codes \cite{horsman_surface_2012}, with parameters $\llbracket  L^2,1,L \rrbracket$ ($L$ being the size of the lattice) for the description of the decoder. The matrix $\mathbf{H}_X$ is the incidence matrix between vertices  and edges, and the matrix $\mathbf{H}_Z$ is the incidence matrix between squares and edges. The two Tanner graphs corresponding to $\mathbf{H}_X$ and $\mathbf{H}_Z$ are isomorphic and they obviously correspond to a lattice as well, as depicted in Fig. \ref{fig:lattice_numbering} for the rotated planar code.
We also consider toric codes, which are again characterized by a $L\times L$ lattice, such that the edges and vertices at the boundaries coincide. The toric code has parameters $\llbracket 2L^2,2,L \rrbracket$.

\section{Decoding of toric and surface codes}
\label{sec:state_of_art}
In this section we summarize the state-of-the-art decoders for toric and surface codes, and motivate our contribution.

The decoding of toric and surface codes is usually carried out using the MWPM algorithm \cite{fowler_minimum_2014}. After measuring the syndrome, a complete graph is constructed, where each vertex represents an unsatisfied check; each edge is assigned a weight, which is the minimum number of qubits separating the two checks connected by the edge; this weight assignment is done using Dijkstra's algorithm \cite{dijkstra_note_2022}. Then, the \textit{blossom algorithm} \cite{edmonds_paths_1965} is run on this graph, with the goal of finding the minimum weight perfect matching.
Since the overall complexity of the MWPM decoder on an $L\times L$ lattice is $\mathcal{O}(L^6\log L)$ \cite{higgott_pymatching_2022}, many efficient implementations have been proposed to reduce this complexity: for instance, in \cite{fowler_minimum_2014} a fully parallel implementation is proposed, able to run with a complexity of $\mathcal{O}(1)$, given a uniform 2-D array of finite speed processing elements and an external memory able to store all the detection events and matching data for the entire duration of a hypothetical computation, while Higgott \textit{et al.} proposed, in \cite{higgott_pymatching_2022}, that each unsatisfied check could be restricted to be matched within a local neighborhood, or that Dijkstra's algorithm could be optimized to avoid an all-to-all search \cite{higgott_sparse_2023}.
The main competitor of MWPM is the UF decoder \cite{delfosse_almost-linear_2021}. This decoding scheme is also divided into two parts: first, the \textit{syndrome validation} process is carried out, where \textit{clusters} are defined to initially correspond to each unsatisfied check, and are then iteratively grown to absorb the nearest check neighbors. A cluster stops growing when it contains an even number of unsatisfied checks, and two clusters merge together if they touch each other. After all the clusters have stopped growing, the \textit{peeling decoder} is run on each cluster.
The complexity of the UF decoder, in its efficient implementation proposed in \cite{delfosse_almost-linear_2021}, is almost linear in the blocklength. 
There are also several alternative approaches to UF, which are able to improve the decoding latency. For instance, in \cite{stm_decoder}, the authors propose a spanning tree (ST) matching decoder which is able to correct up to the code distance. In addition, they also propose a lower complexity decoder, called the \textit{rapid-fire decoder}, which is able to improve the decoding latency at the cost of some performance degradation.

Message passing decoders such as BP are, in principle, not suitable for topological codes, mainly because of the presence of weight-4 symmetric stabilizers that behave as trapping sets for the decoder \cite{raveendran_trapping_2021}; because of this, several techniques have been developed to improve the performance of BP on these codes. One of the most common approaches is to apply the well-known ordered statistics technique after several rounds of BP \cite{panteleev_degenerate_2021}; this decoder is known as BP-OSD and has a complexity of $\mathcal{O}(n^3)$, thus making it infeasible for large-scale implementation.

For the majority of decoders proposed for topological codes, low decoding error probability comes with the price of increased decoding complexity. MWPM, for instance, has a complexity that is cubic in the blocklength and it is known to be infeasible in practice; the efficient implementations available, in particular those of \cite{higgott_pymatching_2022} and \cite{higgott_sparse_2023}, offer a significant speedup that still vanishes for large blocklength.
Another major issue in decoders for topological codes, which is often overlooked, is the hardware implementation aspect of the algorithms. In fact, the vast majority of these decoders turn out to be inefficient when implemented on FPGAs. It is well-known that fixed and predictable access to pre-allocated memory is much faster than random access to irregular, pointer-based dynamic memory; in particular, FPGAs are particularly suitable for the former type of memory access. Moreover, modifications of dynamic structures (like adding nodes to a tree, or modifying its connectivity) are also operations which are known to introduce a high amount of latency, even when parallelized, due to the high overhead required. Another aspect to consider is that the usage of dynamic memories is usually more expensive in terms of energy consumption, as they often require additional overhead for their management, as well as more complex hardware. Also, energy consumption is a critical aspect of any quantum error correction decoder, as it runs on a chip which is embedded in a cryogenic environment with a strict power budget.
MWPM and spanning tree decoders, for example, make use of Dijkstra's algorithm to construct syndrome graphs on which they perform the perfect matching. Dijkstra's algorithm is not desirable for an efficient FPGA implementation, primarily due to its inherently sequential nature and dependence on dynamic data structures (priority queues).

Similarly, the UF decoder relies on dynamic, pointer-based data structures and complex tree manipulations. The two operations, $\mathtt{union()}$ and $\mathtt{find()}$, often involve traversing and modifying tree structures, which require frequent pointer updates and non-uniform memory access patterns.

In this scenario, it seems reasonable to revisit the BF decoder which utilizes extremely simple update rules that make it much easier to implement in hardware such as FPGA. However, because of the low degree of variable nodes in the Tanner graph of the considered quantum codes (toric codes are regular with variable degree of 2, while surface codes also have degree-1 nodes), and because of error degeneracy, we need to make significant modifications to the traditional BF algorithm in order to decode toric and surface codes. We will show that, although these modifications impact on the asymptotic complexity of the decoder, they significantly improve the performance of the decoder comparing to that which uses traditional BF; at the same time, our algorithm does not make use of any dynamic data structures, and only utilizes elementary operations on arrays of fixed length which can be pre-allocated in memory, thus making our decoder particularly suitable for efficient hardware implementation compared to the alternatives. Nevertheless, the goal of this paper is to merely present and explain the algorithm, without presenting an explicit hardware implementation; an optimized hardware implementation, together with a detailed comparison with those of other decoders, requires a specific in-depth study and is therefore left for future work.

\section{The proximity vector}
\label{sec:proximity}

In this section we introduce the \textit{proximity vector}, a vector that the decoder uses as a bit flipping criterion. We describe the rationale behind it, and we describe how it can be efficiently computed while decoding. 

The proximity vector is a heuristic weight assignment to the variable nodes and check nodes in the Tanner graph, and we define it to be the sum of multiple contributions which we call \textit{proximity influences}, which we describe hereafter. Let $c_j$ be an unsatisfied check; we want variable and check nodes neighboring $c_j$ to have the highest weights, while the weights should decrease with increasing distance from $c_j$. A simple way to achieve this is by computing the proximity influence recursively, as follows:
\begin{definition}
   Let $c_j$ be an unsatisfied check, and let all the other check nodes be satisfied. We define $\boldsymbol{\gamma}^{(0)}(c_j)$ to be a $1\times m$ vector such that
    \begin{equation}
        \gamma_{i}^{(0)}(c_j)= \begin{cases}
        1\ \mathrm{for}\ i=j \\
        0\ \mathrm{otherwise}\ ,
    \end{cases}
    \end{equation}
   and let $\boldsymbol{\nu}^{(0)}(c_j)$ be a $1 \times n$ vector such that
   \begin{equation}
   \label{eq:nu_zero}
       \boldsymbol{\nu}^{(0)} = \boldsymbol{\gamma}^{(0)} \cdot \mathbf{H}\ .
   \end{equation}
   Then, let $\boldsymbol{\gamma}^{(\ell)}(c_j)$ and $\boldsymbol{\nu}^{(\ell)}(c_j)$ be defined recursively as
   \begin{equation}
        \label{eq:compute_vectors}
    \begin{cases}
        \boldsymbol{\gamma}^{(\ell)}(c_j) = \boldsymbol{\nu}^{(\ell-1)}(c_j) \cdot \mathbf{H}^T \\
        \boldsymbol{\nu}^{(\ell)}(c_j) = \boldsymbol{\gamma}^{(\ell)}(c_j) \cdot \mathbf{H}\ ,
    \end{cases}
\end{equation}
for $\ell=1,2,...,D$, $D$ being a fixed positive integer. We call $\boldsymbol{\nu}^{(\ell)}(c_j)$ the proximity influence of $c_j$ on qubits (or variable nodes) of depth $\ell$, and $\boldsymbol{\gamma}^{(\ell)}(c_j)$ the proximity influence of $c_j$ on checks of depth $\ell$. 
\end{definition}
We then combine the proximity influence of each unsatisfied check node to obtain the proximity vectors by adding, for each variable and check node, the values of each proximity influence.
\begin{definition}
    Let $c_1,...,c_m$ be all the unsatisfied check nodes, and let $\boldsymbol{\nu}^{(D)}(c_1),...,\boldsymbol{\nu}^{(D)}(c_m)$ and $\boldsymbol{\gamma}^{(D)}(c_1),...,\boldsymbol{\gamma}^{(D)}(c_m)$ be the respective proximity influences on qubits and checks. We define $\boldsymbol{\nu}^{(D)}$ and $\boldsymbol{\gamma}^{(D)}$ to be the proximity vectors on qubits and check nodes of depth $D$, respectively, such that:
\begin{equation}
\label{eq:proxy_sum}
    \begin{cases}
        \boldsymbol{\nu}^{(D)} = \sum_{i=1}^{m} \boldsymbol{\nu}^{(D)}(c_i) \\
        \boldsymbol{\gamma}^{(D)} = \sum_{i=1}^{m} \boldsymbol{\gamma}^{(D)}(c_i)
    \end{cases}.
\end{equation}
It is also possible to define $\boldsymbol{\nu}^{(D)}$ and $\boldsymbol{\gamma}^{(D)}$ by setting $\boldsymbol{\gamma}^{(0)}=\mathbf{s}$, and then using (\ref{eq:nu_zero}) and (\ref{eq:compute_vectors}) directly.
\end{definition}

Since the value of $D$ will be fixed in the rest of the paper, we simply refer to the proximity influences as $\boldsymbol{\nu}(c_j)$ and $\boldsymbol{\gamma}(c_j)$, and to the vectors as $\boldsymbol{\nu}$ and $\boldsymbol{\gamma}$. By construction, the values of the influences $\boldsymbol{\nu}(c_j)$ and $\boldsymbol{\gamma}(c_j)$ are highest for the nearest neighbors of $c_j$, and decrease with increasing distance of a variable (check) node from $c_j$. To see this, it is convenient to write $\boldsymbol{\nu}^{(D)},\boldsymbol{\gamma}^{(D)}$ in the following recursive way. Let $\mathcal{N}(c_j) \subseteq V$ be the neighborhood of check node $c_j$, and $\mathcal{N}(v_j) \subseteq C$ the neighborhood of variable node $v_j$; we define $\mathcal{N}^d(c_j) \subseteq V$ to be the depth-$d$ neighborhood of $c_j$,  \textit{i.e.}, the shortest path between $c_j$ and its depth-$d$ neighbors has length $2d-1$; we can also define $\mathcal{N}^d(v_j) \subseteq C$ in a similar way. For every check node $c_k \in C$ and variable node $v_q \in V$:
\begin{equation}
\label{eq:metric_evolution}
\begin{cases}
        \gamma_k^{(i)}(c_j) = \gamma_k^{(i-1)}(c_j) + \sum_{v_z\in \mathcal{N}(c_k)}\nu_z^{(i-1)}(c_j) \\
        \nu_q^{(i)}(c_j) = \nu_q^{(i-1)}(c_j) + \sum_{c_z\in \mathcal{N}(v_q)}\gamma_z^{(i-1)}(c_j).
\end{cases}
\end{equation}
For $i=0$ we have $\nu_q^{(0)}(c_j)=1$ for all variable nodes $v_q\in\mathcal{N}(c_j)$ and 0 otherwise; for $i=1$ we have $\nu_q^{(1)}(c_j)=1$ for all  variable nodes $v_q\in\mathcal{N}^2(c_j)$; moreover, $\nu_q^{(1)}(c_j)>1$ for all  variable nodes $v_q\in\mathcal{N}(c_j)$. Let $r>i+1$ and $m<i+1$; in general we have $\nu_q^{(i)}(c_j)=0$ for all variable nodes in $\mathcal{N}^r(c_j)$, $\nu_q^{(i)}(c_j)=1$ for all variable nodes in $\mathcal{N}^{i+1}(c_j)$, and  $\nu_q^{(i)}(c_j)>1$ for variable nodes in $\mathcal{N}^m(c_j)$. In other words, the value of the proximity influence for each node at distance $2d-1$ from $c_j$ remains $0$ until the $d+1$ iteration, and then increases according to (\ref{eq:metric_evolution}). This means that, fixing the number of iterations, the closer is a node to $c_j$, the higher is the proximity influence for that node. A similar argument can be made for $\boldsymbol{\gamma}(c_j)$.

The proximity vector on qubits $\boldsymbol{\nu}$ and on checks $\boldsymbol{\gamma}$ is the natural superposition of all of the influences exerted by all of the unsatisfied checks. As a result, variable and check nodes which are more distant from unsatisfied nodes (we say more \textit{isolated}) will have a lower weight assigned, while variable and check nodes that are near to many unsatisfied checks will have higher weights assigned. Note that this definition of proximity vector is heuristic, and alternative definitions can be provided to capture the individual influence of a check node. Our definition of proximity vector is convenient for the reason that it only involves integer values, thus it can be stored using a fixed (and low) number of bits.

\subsubsection{Efficient computation of the proximity vector}
\label{sec:lowcomplexity}
In our decoder, it is essential to update the proximity vectors on checks and qubits after each flip: in other words, after a bit is flipped, its neighboring checks will be satisfied\footnote{Because of the nature of our decoder, each bit flip never generates new unsatisfied checks.}, and their influence should be removed from the proximity vector. Specifically, if the flipping of variable node $v_i$ causes its two neighboring check nodes $c_j$ and $c_k$ to become satisfied, the updated proximity vectors shall be 
\begin{equation}
\label{eq:proxy_subtract}
\begin{cases}
    \boldsymbol{\gamma}' = \boldsymbol{\gamma}- (\boldsymbol{\gamma}(c_j) + \boldsymbol{\gamma}(c_k))\\
    \boldsymbol{\nu}' = \boldsymbol{\nu} - (\boldsymbol{\nu}(c_j) + \boldsymbol{\nu}(c_k)).\\
\end{cases}
\end{equation}
To take into account this update of the vector after each flip, one could in principle recompute (\ref{eq:proxy_sum}), setting $\boldsymbol{\gamma}^{(0)}=\mathbf{s}'$, where $\mathbf{s}'$ is the updated syndrome after the bit flip; however, this would negatively impact the complexity of the algorithm, as each bit flip would require an additional $\ell(nd_v + md_c)$ operations (multiplication of sparse matrices).

\begin{figure}[!htb]
    \begin{subfigure}{0.5\textwidth}
    \centering
    \input{tikz_figures/rotated_influece.tikz}
    \caption{The depth-1 influence of $c_8$ (highlighted in red) is computed on a $4\times 5$ surface code. The node label of $c_8$ corresponds to the node $c_0$ in the $L=3$ rotated code.}
    \label{fig:lattice_influence}
\end{subfigure}

\begin{subfigure}{0.5\textwidth}
    \centering
    \input{tikz_figures/rotated_influence_2.tikz}
    \caption{The influence of Fig. \ref{fig:lattice_influence} is shifted to the node $c_{15}$ on the $4\times 5$ surface code: in this case, $\sigma_x=1$ and $\sigma_y=1$. Then, the values of the shifted metrics are assigned to the nodes of the $L=3$ rotated code according to the illustrated mapping between labels. }
    \label{fig:lattice_influence_shift}
\end{subfigure}

\caption{Example of shifting of the proximity influence. The labeling on the variable (black) and check (red) nodes inside the parentheses is the labeling for the $L=3$ rotated code, while the labeling outside the parentheses is the one for the $4\times 5$ surface code. In this example, we compute the influence of $c_{15(2)}$ by shifting the influence of $c_{8(0)}$. The nodes colored in gray are the ones with a non-zero proximity metric; notice how some of the values are discarded after shifting, i.e. the values of $c_9$ and $c_{16}$.}
\label{fig:lattice_influence_tot}
\end{figure}
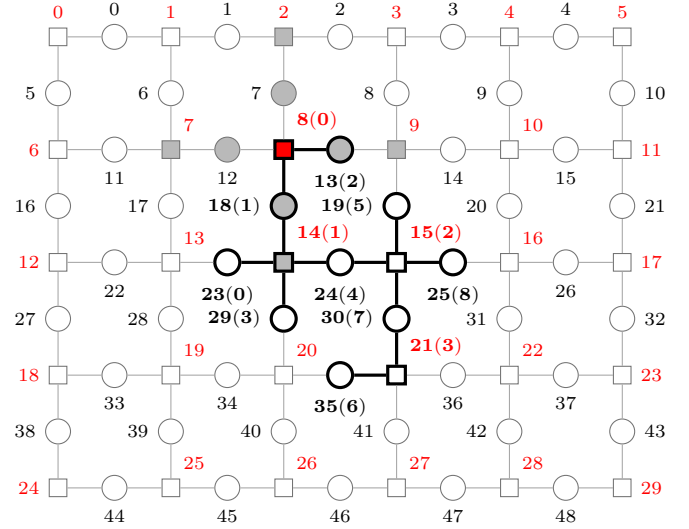
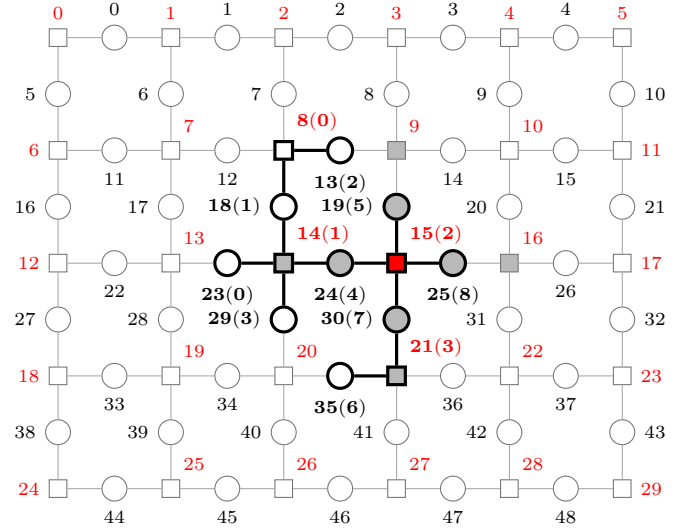

Instead, we exploit the labeling we have assigned to variable and check nodes in the surface code. The idea is to pre-compute the proximity influence of an arbitrary check node over a larger surface code, say $\boldsymbol{\gamma}(c_1),\boldsymbol{\nu}(c_1)$, store it prior to the decoding process, and then compute online $\boldsymbol{\gamma}(c_i),\boldsymbol{\nu}(c_i)$ when needed by appropriately permuting $\boldsymbol{\gamma}(c_1)$ and $\boldsymbol{\nu}(c_1)$. We illustrate this process for the rotated surface code, although it can be extended for other types of surface codes. Assume that we label the variable nodes with non-negative integers row-wise in increasing order and we do the same for the check nodes; an example is provided in Fig. \ref{fig:lattice_numbering} for a rotated code with $L=3$.  Consider the Tanner graphs $\mathcal{G}_1$ of a rotated code with $L=L_1$, and $\mathcal{G}_2$ of an $L_2'\times L_2''$ surface code , such that $L_2>L_1$ and $L_2''>L_1$.
Let $\mathcal{L}^{(1)}_v, \mathcal{L}^{(1)}_c \in \mathbb{N}^{L_1^2}$ be the labels of the variable and check nodes of $\mathcal{G}_1$, and $\mathcal{L}^{(2)}_v, \mathcal{L}^{(2)}_c  \in \mathbb{N}^{L_2^2}$ be the labels of the variable and check nodes of $\mathcal{G}_2$, respectively.
There are two injective mappings $\phi_c : \mathcal{L}^{(1)}_c \rightarrow \mathcal{L}^{(2)}_c$ and  $\phi_v : \mathcal{L}^{(1)}_v \rightarrow \mathcal{L}^{(2)}_v$ from the labels of the check and variable nodes of $\mathcal{G}_1$ to the labels of the check and variable nodes of $\mathcal{G}_2$. Let $i,j \in \mathcal{L}^{(1)}_c$.
Assume that we have calculated $\boldsymbol{\nu}(c_{\phi_c(i)})$ and $\boldsymbol{\gamma}(c_{\phi_c(i)})$ for some arbitrary check node $c_i$ using (\ref{eq:compute_vectors}) on $\mathcal{G}_2$, and that we now wish to compute $\boldsymbol{\nu}(c_j)$ and $\boldsymbol{\gamma}(c_j)$, for some $j \neq i$, by appropriately permuting $\boldsymbol{\nu}(c_{\phi_c(i)})$ and $\boldsymbol{\gamma}(c_{\phi_c(i)})$. Let $\rho_c$ be the number of check in each row of $\mathcal{G}_2$ and $\rho_{vo},\rho_{ve}$ be the number of variable nodes in even and odd rows of $\mathcal{G}_2$, respectively.  By exploiting the indexing we have defined for the check and variable nodes, we can define a \textit{vertical} shift $\sigma_y$ and an \textit{horizontal} shift $\sigma_x$:
\begin{equation}
\label{eq:shift}
    \begin{cases}
        \sigma_y = \lfloor \frac{\phi_c(j)}{\rho_c}\rfloor - \lfloor \frac{\phi_c(i)}{\rho_c}\rfloor, \\
        \sigma_x = \phi_c(j) \mod \rho_c\ -\ \phi_c(i) \mod \rho_c\ , \\
    \end{cases}
\end{equation}
Note that $\sigma_x$ and $\sigma_y$ can also assume negative values. Depending on the surface code we are considering, it is possible to express the index transformation in closed form, using $\sigma_x$ and $\sigma_y$. 
Let $k_c,k'_c \in \mathcal{L}_c^{(2)}$ and $k_v,k'_v \in \mathcal{L}_v^{(2)}$. We can express a coordinate shift using two linear maps from $k_c$ to $k'_c$ and from $k_v$ to $k'_v$ respectively, such that
\begin{equation}
\begin{cases}
        k'_c = k_c + \sigma_x + \sigma_y \rho_c\\
        k'_v = k_v +\sigma_x + \sigma_y (\rho_{vo}+\rho_{ve}).
\end{cases}
\label{eq:coord_shift}
\end{equation}
Thus, we have  $\nu_{k'_v}(c_{\phi_c(j)})= \nu_{k_v}(c_{\phi_c(i)})$ and $\gamma_{k'_c}(c_{\phi_c(j)}) = \gamma_{k_c}(c_{\phi_c(i)})$, for all $k'_c$ and $k'_v$; finally, we apply the inverse mappings $\phi_c^{-1}$, $\phi_v^{-1}$ to all the elements in the codomain of these functions, such that $\nu_{q}(c_j) = \nu_{\phi_v(q)}(c_{\phi_c(j)})$ and $\gamma_{p}(c_j) = \gamma_{\phi_c(p)}(c_{\phi_c(j)})$ for any $q \in \mathcal{L}_v^{(1)}$ and $p \in \mathcal{L}_c^{(1)}$. This procedure is illustrated in Algorithm \ref{alg:shift}, where $j \in \mathcal{L}^{(1)}_c$ is the label of the check nodes of which we want to compute the proximity vectors. An example of the application of Algorithm \ref{alg:shift} is illustrated in Fig. \ref{fig:lattice_influence_tot}. Specifically, we pick $\mathcal{G}_1$ to be a $L_1=3$ rotated code and $\mathcal{G}_2$ to be a $4\times 5$ surface code. The example illustrates how to shift the proximity vectors from $c_{\phi_c(0)=8}$ to $c_{\phi_c(2)=15}$. In Fig. \ref{fig:lattice_influence} the proximity metrics $\boldsymbol{\gamma}(c_{\phi_c(0)})$ and  $\boldsymbol{\nu}(c_{\phi_c(0)})$ are assumed to be pre-computed; in Fig. \ref{fig:lattice_influence_shift} we want to compute $\boldsymbol{\gamma}(c_2)$ and  $\boldsymbol{\nu}(c_2)$, which is mapped to $\boldsymbol{\gamma}(c_{\phi_c(2)=15})$ and  $\boldsymbol{\nu}(c_{\phi_c(2)=15})$. We compute $\sigma_x=1$ and $\sigma_y=1$ and apply (\ref{eq:coord_shift}) to each variable and check node. The reader can verify that, for instance, check node 7 is mapped to check node 14, and thus to check node $\phi^{-1}_c(14)=1$ in the $L=3$ surface code. On the other hand, check node 2 is mapped to check node 9, which doesn't belong to the image of $\phi_c$, and thus $\boldsymbol{\gamma}(c_{\phi^{-1}(c_9)})=\emptyset$. The values for $L_2',L_2''$ are such that the mapping in (\ref{eq:shift}) gives the correct result for every $\sigma_x,\sigma_y$, which happens for $L_2',L_2''>L+D$.

\begin{algorithm}
    \caption{$\mathtt{Shift\_influence}$}
     \label{alg:shift}
  \textbf{Input}:  $j$ \\
    \textbf{Global}: $\boldsymbol{\gamma}(c_{\phi_c(1)})$,  $\boldsymbol{\nu}(c_{\phi_c(1)})$, $L_2$, $\phi_v$, $\phi_c$ \\
  \textbf{Output}: $\boldsymbol{\gamma}(c_j)$,  $\boldsymbol{\nu}(c_j)$
  \input{Algorithms/alg_shift_influence}
\end{algorithm}

\subsubsection{Auxiliary proximity influence}
\label{sec:match_influence}
We also define an auxiliary proximity influence of a check node $c_j$, that we denote as  $\boldsymbol{\alpha}(c_j)$.
\begin{definition}
   The auxiliary proximity influence $\boldsymbol{\alpha}(c_j)$ is a $1 \times n$ vector such that $\alpha_i(c_j)$ is the length of the shortest path between the variable node $v_i$ and the check node $c_j$. For example, if a variable node $v_i$ is directly connected to $c_j$, then $\alpha_i(c_j)=1$.
\end{definition}
Note that the proximity influences $\boldsymbol{\nu}(c_j)$ and $\boldsymbol{\alpha}(c_j)$ share the same non-zero positions, but in general they have different values: while $\boldsymbol{\nu}(c_j)$ will have higher values for the variable nodes close to $c_j$, in $\boldsymbol{\alpha}(c_j)$ the variable nodes connected to $c_j$ will have value $1$, those at distance 2 will have value $2$, and so on. We can compute $\boldsymbol{\alpha}(c_j)$ in a similar way to $\boldsymbol{\nu}(c_j)$. Thus, it is sufficient to construct $\boldsymbol{\alpha}(c_1)$ offline, and apply Algorithm~\ref{alg:shift} to create $\boldsymbol{\alpha}(c_j)$, for any $j$. We illustrate in Section \ref{sec:matching} how we make use of the auxiliary proximity influence to correct errors.
Finally, we define two subroutines, illustrated in Algorithm~\ref{alg:shift_add} and Algorithm~\ref{alg:shift_remove}. The $\mathtt{Compute\_proximity\_vector}$ subroutine takes as input the syndrome $\mathbf{s}$ and generates the proximity vectors $\boldsymbol{\nu}$ and $\boldsymbol{\gamma}$, using Algorithm \ref{alg:shift} to compute individual influences and summing them up according to (\ref{eq:proxy_sum}). The $\mathtt{Shift\_and\_remove}$ subroutine takes as input the residual syndrome $\mathbf{s}'$, as well as the proximity vectors $\boldsymbol{\nu}$ and $\boldsymbol{\gamma}$, and updates them using  (\ref{eq:proxy_subtract}). 

\begin{algorithm}
    \caption{$\mathtt{ Compute\_proximity\_vector}$}
     \label{alg:shift_add}
  \textbf{Input}: $\mathbf{s}$\\
    \textbf{Global}:  $\boldsymbol{\gamma}(c_1)$,  $\boldsymbol{\nu}(c_1)$, $L$ \\
  \textbf{Output}: $\boldsymbol{\nu}$,  $\boldsymbol{\gamma}$
    \input{Algorithms/alg_compute_proximity}
\end{algorithm}

\begin{algorithm}
    \caption{$\mathtt{Shift\_and\_remove}$}
     \label{alg:shift_remove}
  \textbf{Input}:  $\boldsymbol{\nu}$,  $\boldsymbol{\gamma}$,  $\mathbf{s}'$\\
    \textbf{Global}:  $\boldsymbol{\gamma}(c_1)$,  $\boldsymbol{\nu}(c_1)$, $L$ \\
  \textbf{Output}: $\boldsymbol{\nu}'$,  $\boldsymbol{\gamma}'$
    \input{Algorithms/alg_shift_and_remove}
\end{algorithm}

\section{Progressive-Proximity Bit-Flipping}
\label{sec:decoder}
We are now ready to describe our proposed decoder, which we call \textit{Progressive-Proximity Bit-Flipping} (PPBF). It is illustrated in Algorithm \ref{alg:bitmatching}, and it is composed of two decoding steps: the first is called \textit{Preliminary BF}, and the second is called \textit{Iterative matching}; these two algorithms are illustrated in Algorithm \ref{alg:pBF} and Algorithm \ref{alg:iter-matching}, respectively.

\begin{algorithm}
    \caption{Progressive-Proximity Bit-Flipping}
    \label{alg:bitmatching}
    \textbf{Input}: $\mathbf{s}$, $\mathbf{H}$\\
    \textbf{Output}: $\hat{\mathbf{e}}$
    \input{Algorithms/alg_bit_matching}
\end{algorithm}

\subsection{Preliminary BF}
In the preliminary BF step of Algorithm \ref{alg:pBF}, the vector $\mathbf{u}$ containing the number of unsatisfied checks for each variable node is first computed (line~\ref{alg1:counters}), and only variable nodes involved in two unsatisfied checks will be considered for flipping. Among all the variable nodes $i$ such that $u_i=2$, we flip the one which achieves the minimum value of the proximity vector $\nu_i'$ (lines \ref{alg1:minproxy}-\ref{alg1:flip}); then, the proximity vector is updated using Algorithm \ref{alg:shift_remove}, such that the influence of the checks satisfied after the flip is removed from $\boldsymbol{\nu}'$ (line \ref{alg1:shiftremove}). The residual syndrome is then computed (line \ref{alg1:residual}) and another iteration is performed, until there are no variable nodes $i$ such that $u_i=2$.

\begin{algorithm}
    \caption{$\mathtt{ Preliminary\_BF}$}
     \label{alg:pBF}
  \textbf{Input}: $\mathbf{s}$, $\mathbf{H}$, $\boldsymbol{\nu}$

  \textbf{Output}: $\hat{\mathbf{e}}$, $\hat{\mathbf{s}}$, $\boldsymbol{\nu}'$
    \input{Algorithms/alg_preliminary_bf}
\end{algorithm}

\subsection{The iterative-matching routine}
\label{sec:matching}

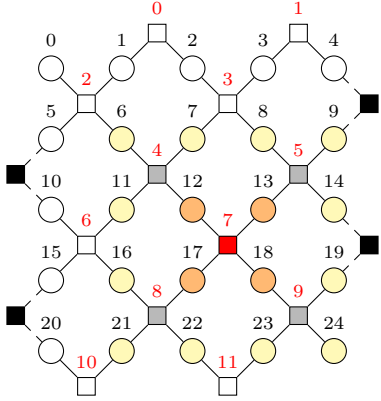
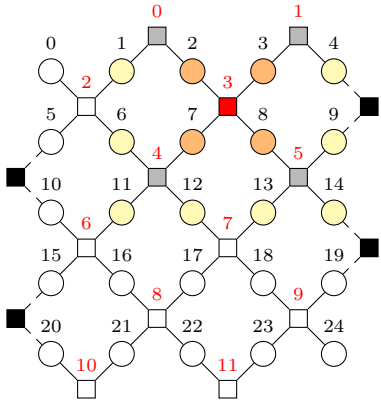
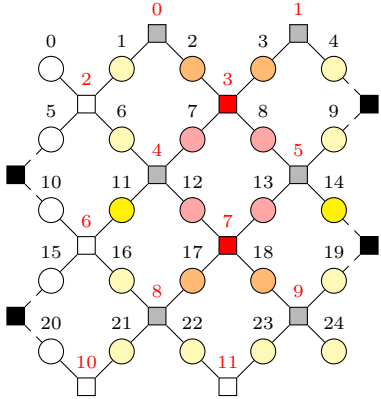
\begin{figure}
    \begin{subfigure}{0.5\textwidth}
    \centering
    \input{tikz_figures/matching_rotated_1.tikz}
    \caption{Proximity influence of depth 2 for the check node $c_7$; different colorings of variable nodes correspond to their distance from $c_7$: orange nodes are at distance 1, yellow at distance 2.}
    \label{fig:matching_influence_1}
\end{subfigure}

\begin{subfigure}{0.5\textwidth}
    \centering
    \input{tikz_figures/matching_rotated_2.tikz}
    \caption{Proximity influence of depth 2 for $c_3$.}
    \label{fig:matching_influence_2}
\end{subfigure}

\begin{subfigure}{0.5\textwidth}
    \centering
    \input{tikz_figures/matching_rotated_3.tikz}
    \caption{Combination of the previous two proximity influences: variable nodes which were assigned yellow and orange are now colored red; those nodes which were assigned yellow and yellow are now deep yellow. Finally, the variable nodes which were assigned orange/white or yellow/white have maintained their color. The set of red variable nodes coincide with the union of all the possible error patterns that satisfy the matching, \textit{\textit{i.e.}}, $\{v_{7},v_{12}\}$ and $\{v_{8},v_{13}\}$.}
    \label{fig:matching_influence_3}
    
\end{subfigure}
\caption{Example of the usage of the auxiliary proximity influence of checks in Algorithm \ref{alg:iter-matching}. In the example, $c_3$ and $c_7$ are unsatisfied checks, and the decoder has to find the shortest path between them, assuming it starts with $c_7$ as pivot.  }
\label{fig:combine_influence}
\end{figure}

Here we present the iterative matching routine, which is specified in Algorithm \ref{alg:iter-matching}. A matching decoder is a decoding algorithm specifically tailored for surface codes, which aims to pair together couples of unsatisfied checks, estimating the qubits on the shortest path connecting them as being in error. This comes from the fact that every error on the surface code can be interpreted as a path on the lattice, and the two endpoints of this path are the two unsatisfied checks generated by the error; however, errors occurring on boundary qubits only generate one unsatisfied check. To address this issue, it is common to add additional ``dummy'' unsatisfied check nodes to the boundaries of the surface code, as proposed in \cite{fowler_minimum_2014} (notice that this is not necessary for the toric code). These checks are illustrated in black in Fig. \ref{fig:combine_influence}. We utilize the proximity vector $\boldsymbol{\gamma}$ to match pairs of unsatisfied checks together; specifically, we start by identifying the unsatisfied check $c_i$ with the lowest proximity vector entry $\gamma_i$ (line 4) calling it a \textit{pivot node}, and compute its auxiliary proximity influence $\boldsymbol{\alpha}(c_i)$. Among all the other unsatisfied checks, we pick the one at the smallest distance from the pivot (if there is more than one candidate at the same distance, we choose the check $c_j$ with lowest proximity vector entry $\gamma_j$); we call this the \textit{target node} (line \ref{alg2:target}), 
and denote its distance from the pivot by $\delta$. After computing the auxiliary vector $\boldsymbol{\alpha}(c_j)$, we compute $\boldsymbol{\alpha}(c_i)+\boldsymbol{\alpha}(c_j)$; the result of this operation is a vector such that if the $k$-th element is equal to $\delta+1$, the variable node $v_k$ belongs to the shortest path between $c_i$ and $c_j$. If the number of entries of $\boldsymbol{\alpha}(c_i)+\boldsymbol{\alpha}(c_j)$ that are equal to $\delta+1$ is equal to $\delta$, it means that there is only one shortest path between $c_i$ and $c_j$, that corresponds to the most likely error matching that syndrome; on the other hand, if the number of entries of $\boldsymbol{\alpha}(c_i)+\boldsymbol{\alpha}(c_j)$ that are equal to $\delta+1$ is larger than $\delta$, it means that the corresponding error is degenerate, as there is more than one shortest path between $c_i$ and $c_j$. An example of this is illustrated in Fig. \ref{fig:combine_influence}, where there are two possible paths of length 2 between $c_3$ and $c_7$, and the number of variable nodes such that $\boldsymbol{\alpha}(c_3)+\boldsymbol{\alpha}(c_7)=3$ is more than 2. In the first case, it is sufficient to flip all the variable nodes associated with the value $\delta+1$, while in the latter case one of the possible degenerate errors must be chosen. We distinguish these two scenarios in line \ref{alg2:choice}. To pick one among the possible paths, we compute the auxiliary influence of a third check node $c_z$, of position such that it is horizontally aligned with $c_i$ and vertically aligned with $c_j$; lines 16-22 are dedicated to computing the position $z$. In particular, we define $\sigma_x$ to be the \textit{horizontal shift} between $c_i$ and $c_j$ and $\sigma_y$ to be the \textit{vertical shift} between $c_i$ and $c_j$. Once $\boldsymbol{\alpha}(c_z)$ is computed in lines 22-23, it is easy to check that the variable nodes $v_k$ such that $\alpha_k(c_i)+\alpha_k(c_z)=\sigma_x+1$ are those connecting $c_i$ and $c_z$, and those such that $\alpha_k(c_j)+\alpha_k(c_z)=\sigma_y+1$ are those connecting $c_j$ and $c_z$, thus we can flip all of them simultaneously in line 26.

\begin{figure}[!htb]
    \centering
    \input{tikz_figures/matching_rotated_4.tikz}
    \caption{Addition of the extra check as described in Section \ref{sec:matching}. The blue variable nodes are the depth-1 influence of the extra check, and the darker ones are those which will be flipped. In the example, $\sigma_x=\sigma_y=1$. }
    \label{fig:extra_check}
\end{figure}
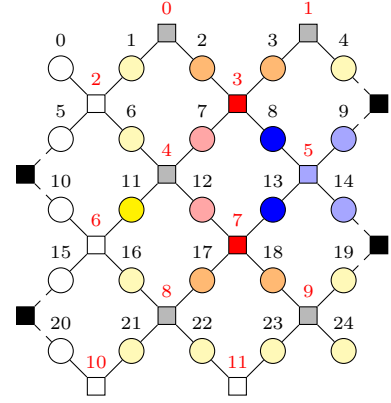

\begin{algorithm}
    \caption{$\mathtt{Iterative\_matching}$}
     \label{alg:iter-matching}
  \textbf{Input}: $\hat{\mathbf{e}}$, $\mathbf{s}$, $\mathbf{H}$, $\boldsymbol{\gamma}$\\
  \textbf{Global}: $\phi_c,\phi_v$ \\
  \textbf{Output}: $\hat{\mathbf{e}}$
 \input{Algorithms/alg_iterative_matching_v2}
\end{algorithm}

\subsection{Decoding of toric codes}
\label{sec:non-toric}
In the case of toric codes it is possible to simplify the computation of the proximity vectors. Consider a $\llbracket 2L^2,2,L \rrbracket$ toric code. The proximity vector of an arbitrary check node is pre-computed on the same code (there is no need to use a larger code), therefore for each node label $i$ we have $\phi(i) = i$; to perform the permutation from a node $i$ to a node $j$, $\sigma_x$ and $\sigma_y$ are computed using (\ref{eq:shift}), and for each $k_c,k_v,k'_c,k'_v \in [1,2L^2]$ we apply
\begin{equation}
        \begin{cases}
        k'_c = \bigl\{ \{k_c+\sigma_x\}\mod L\  + \\ \hspace{25pt} L\lfloor (k_c-1)/L\rfloor +\sigma_y L \bigr\} \mod L^2  \\
         k'_v = \bigl\{\{k_v+\sigma_x\}\mod L\ + \\ \hspace{25pt} L\lfloor (k_v-1)/L\rfloor +2\sigma_y L\bigr\} \mod 2L^2 \\       
    \end{cases}.
\end{equation}
To explain this procedure, we note that the toric code is composed of $L$ rows of $L$ check nodes, for a total of $L^2$ checks, and $2L$ rows of $L$ variable nodes, for a total of $2L^2$ variable nodes. The labels in each row are shifted cyclically modulo $L$ by the term $k_c+\sigma_x \mod L$, which returns a value in $[0,L-1]$. Adding $L\lfloor (k_c-1)/L \rfloor$ returns a value in the same row of $k_c$ but shifted by $\sigma_x$. Finally, a shift of $\sigma_yL$ modulo $L^2$ is applied to get the final label which will be in the same column but a different row. A similar reasoning is applied for $k_v$. The rest of the algorithm remains the same.

\section{Complexity analysis}
\label{sec:complexity}
In this section, we analyze the complexity of the proposed PPBF decoder. The computation of the proximity vector for $c_1$ is done offline and only once, thus it does not contribute to the computational complexity of the algorithm. 
\subsection{Complexity of Algorithms \ref{alg:shift},   \ref{alg:shift_add} and \ref{alg:shift_remove}}
Algorithms \ref{alg:shift},  \ref{alg:shift_add} and \ref{alg:shift_remove} are extensively used in every decoding iteration, therefore it is crucial that their computational complexity is low. Algorithm \ref{alg:shift} simply applies (\ref{eq:shift}) and (\ref{eq:coord_shift}) to shift the proximity influences; this can be done with complexity $\mathcal{O}(1)$. Algorithm \ref{alg:shift_remove}  applies Algorithm \ref{alg:shift} to update the proximity vector; since we update the proximity vector after every flip, Algorithm \ref{alg:shift} is applied twice (every flip satisfies two check nodes) for every Algorithm \ref{alg:shift_remove} call. Thus, Algorithm \ref{alg:shift_remove} has the same order of complexity as Algorithm \ref{alg:shift}. Finally, since Algorithm \ref{alg:shift_add} is applied just once offline to compute $\boldsymbol{\gamma}(c_1)$ and $\boldsymbol{\nu}(c_1)$, it does not contribute to the decoding complexity; in any case, similarly for Algorithm \ref{alg:shift_remove}, it only involves calling Algorithm \ref{alg:shift} $|\mathbf{s}|$ times, thus it has the same complexity. We then conclude that the computational complexity of Algorithms \ref{alg:shift}, \ref{alg:shift_add}  and \ref{alg:shift_remove} is $\mathcal{O}(1)$. We also want to stress that, since the entries of the proximity vectors are integers with bounded values, and since Algorithms \ref{alg:shift}, \ref{alg:shift_add}  and \ref{alg:shift_remove} only involve elementary operations such as integer summation, the memory usage and the runtime of the algorithm will generally be low compared to those of other algorithms with similar computational complexity, but that use floating point calculations.

\subsection{BF decoding complexity}
Assuming that comparing elements of $\mathbf{u}$ and $\boldsymbol{\nu}$ has a negligible impact on the complexity, the only meaningful contribution comes from Algorithm \ref{alg:shift_remove} and the computation of $\mathbf{u}$ (lines 10 and 12 of Algorithm \ref{alg:pBF}), which both have complexity $\mathcal{O}(n)$.

\subsection{Iterative matching complexity}
In each iteration of Algorithm \ref{alg:iter-matching}, all of the operations can be performed in $\mathcal{O}(1)$, except for the $\mathtt{argmin}$ function which is performed in $\mathcal{O}(n)$; since the procedure is repeated for each pair of unsatisfied checks, namely $|\mathbf{s}|/2$ times, making the complexity of the algorithm proportional to $\mathcal{O}(n|\mathbf{s}|)$, and since $|\mathbf{s}| = \mathcal{O}(m)$ \cite{higgott_pymatching_2022}, $m$ being the number of check nodes, and since $m=\mathcal{O}(n)$ (for instance, for the toric code we have $m=n/2$) we have that the complexity of our decoder is $\mathcal{O}(n^2)$.
The MWPM complexity is in the order of $\mathcal{O}(n^{12}\log n^2)$, but efficient implementations are available that achieve very small runtime for relatively small values of $n$ \cite{higgott_pymatching_2022}. The Union Find decoder \cite{delfosse_almost-linear_2021} achieves an almost-linear complexity of $\mathcal{O}(\alpha(n)n)$, where $\alpha(n)$ is the inverse of Ackermann's function \cite{tarjan_efficiency_1975}, which has been shown to be linear for practical values of $n$.


\section{Hardware implementation aspects}
\label{sec:hardware}
The major advantage of our algorithm in comparison to UF, MWPM and ST is that it does not require the storage of data structures that require dynamic memory allocation.  In this paragraph, we first analyze the UF, MWPM and ST decoders in terms of their usage of dynamic memory, then we show how PPBF, in contrast, only requires static access to pre-allocated memory, thus providing advantages in terms of latency, hardware usage, and energy consumption.

The UF algorithm uses tree structures to represent clusters. Each tree is expanded at each iteration, and eventually trees are merged together. The functions that operate on trees in the UF algorithm are called $\mathtt{union()}$ and $\mathtt{find()}$. The function $\mathtt{find()}$ can be decomposed in three fundamental operations: $i)$ \textit{traversal}, where the function navigates through the nodes of the tree, $ii)$ \textit{comparison}, where the function checks if the node is the root or not, and $iii)$ \textit{path compression}, where the structure of the tree is modified such that each leaf points directly to the root.  The function $\mathtt{union()}$ can be decomposed into the operations of $i)$ \textit{find}, where the function applies $\mathtt{find()}$ to a pair of nodes to find their respective root, $ii)$ \textit{comparison}, where the function compares the size of two trees (the smallest is embedded in the largest), $iii)$ \textit{linking}, where the function updates the pointers of the nodes to merge the trees, and $iv)$ \textit{size update}, where the size of the new tree is updated. We also highlight that the size and number of trees that need to be stored is not known \textit{a priori}, and the access to each tree is frequent and unpredictable.
On the other hand, MWPM and ST both make use of Dijkstra's algorithm, which also makes use of trees as data structures. At each iteration, the node at the shortest distance is extracted from the tree, and the tree is updated after the node is extracted. This operation has a similar nature to the \textit{traversal} and \textit{path compression} steps of UF. Since at each iteration one single node is extracted, and the tree consequently updated, Dijkstra's algorithm is intrinsically sequential and thus hard to make parallel. Moreover, the adjacency matrix of the syndrome graph created with Dijkstra's algorithm has unknown dimension, and thus requires dynamic memory allocation for storage.
In comparison, PPBF only requires the storage of $\boldsymbol{\gamma}(c_1)$ and $\boldsymbol{\nu}(c_1)$, which are two integer vectors of length $n$, and the value $L$ which is an integer. Moreover, the knowledge of the map $\phi$ is required, which it is represented by two length-$n$ vectors. The algorithm does not make use of dynamic data structures, but only uses arrays and integers of fixed dimension which can be pre-allocated in memory, increasing the hardware efficiency. The algorithm has a fixed and predictable access to the arrays, indeed all the functions which act on arrays are element-wise operations, which can be made parallel (except for the $\mathrm{arg\ min}$ function). Finally, the proximity vectors can be normalized and stored with finite precision. For these reasons, our algorithm possesses unique advantages for hardware implementation compared to the competing approaches. In Table \ref{tab:summary} we summarize our analysis. UF and MWPM/ST decoder make use of pointer-based data structures, such as trees, which are accessed in an irregular manner and also modified by the algorithm, thus requiring dynamic memory allocation. In comparison, PPBF only makes use of static data structures as arrays, which can be pre-allocated, and that are modified element-wise only.

\begin{table}[]
    \centering
    \begin{tabular}{|c|c|c|c|}
    \hline 
    Decoder & Memory alloc. & Data type & Access to data \\
    \hline
       UF  & Dynamic & Pointer-based & Irregular \\
       \hline 
        MWPM/ST & Dynamic & Pointer-based & Irregular \\
        \hline
        PPBF & Static & Array & Element-wise \\
        \hline 
    \end{tabular}
    \caption{Summary of the memory usage of the analyzed decoders.}
    \label{tab:summary}
\end{table}

\section{Results}
\label{sec:results}
\begin{figure}
    \centering
    \includegraphics[scale=0.65]{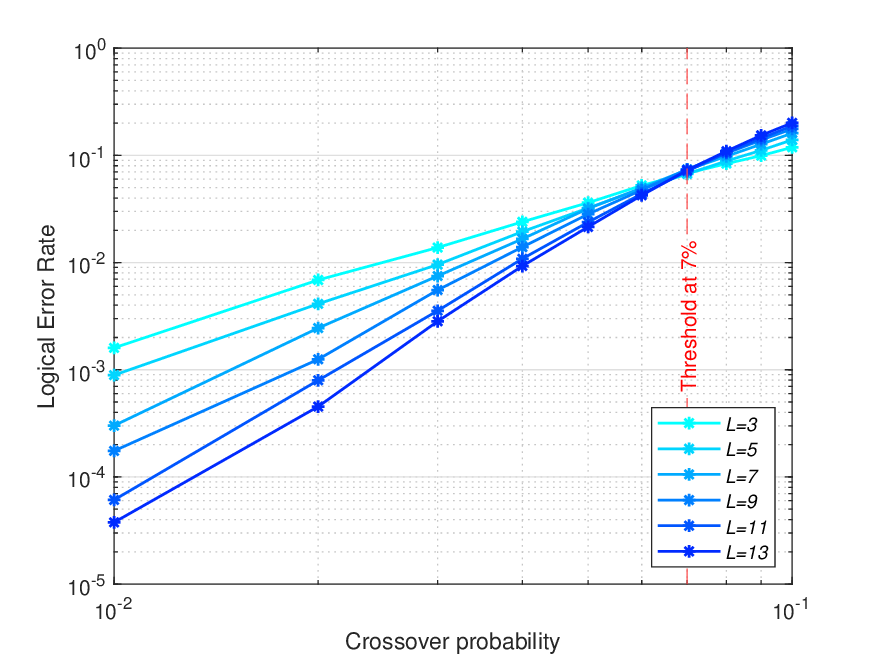}
    \caption{Performance of our decoder on planar rotated codes of different sizes, assuming a BSC. The threshold, which is around $7\%$, is highlighted.}
    \label{fig:rotated}
\end{figure}

\begin{figure}
    \centering
    \includegraphics[scale=0.65]{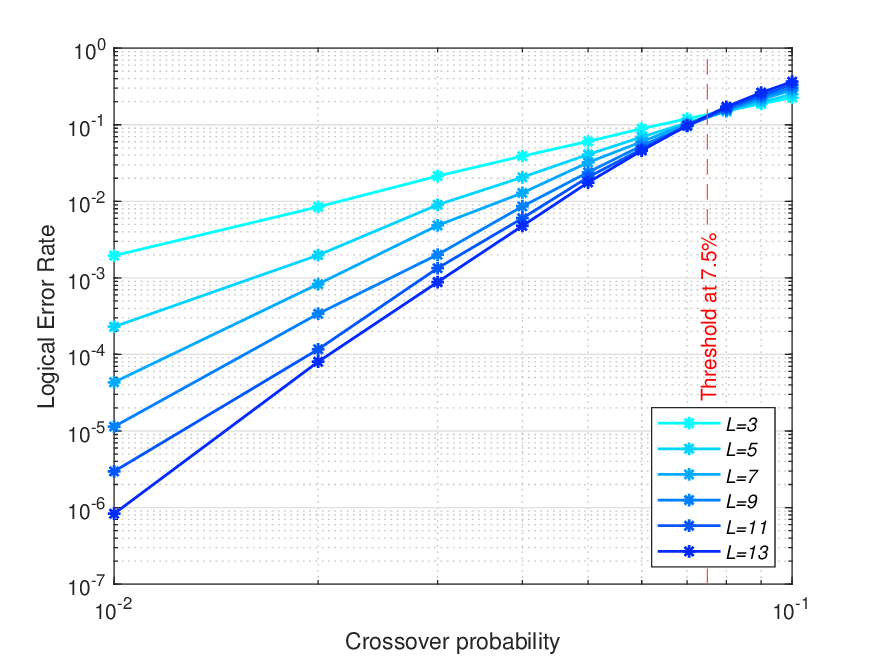}
    \caption{Performance of the proposed decoder on toric codes of different sizes, assuming a BSC. The threshold, which is around $7.5\%$, is highlighted.}
    \label{fig:toric}
\end{figure}

\begin{figure}
\centering
    \includegraphics[scale=0.65]{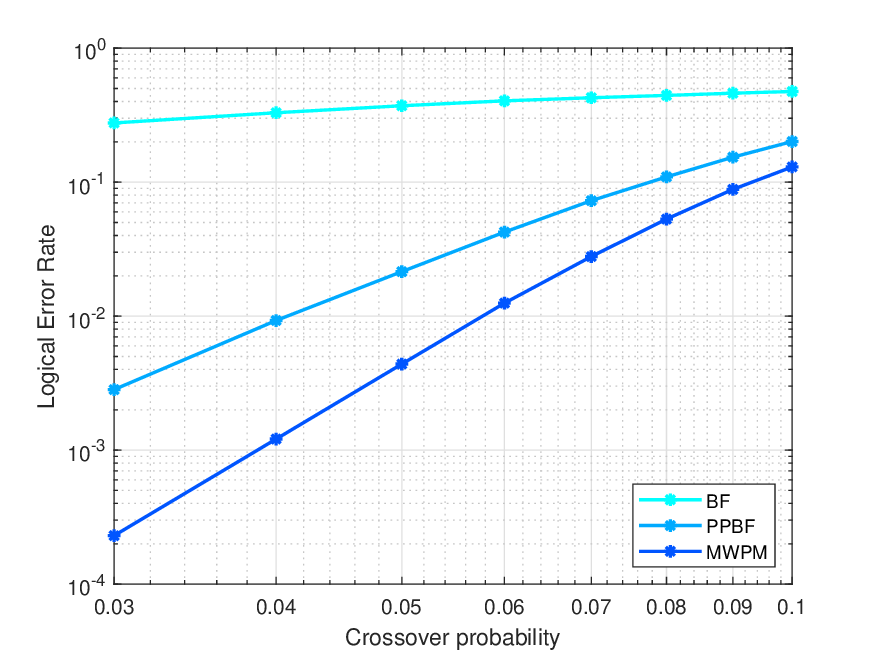}
\caption{Comparison of BF, PPBF and MWPM for the distance 13 rotated code.}
\label{fig:comp_rotated}
\end{figure}

\begin{figure}
\centering
    \includegraphics[scale=0.65]{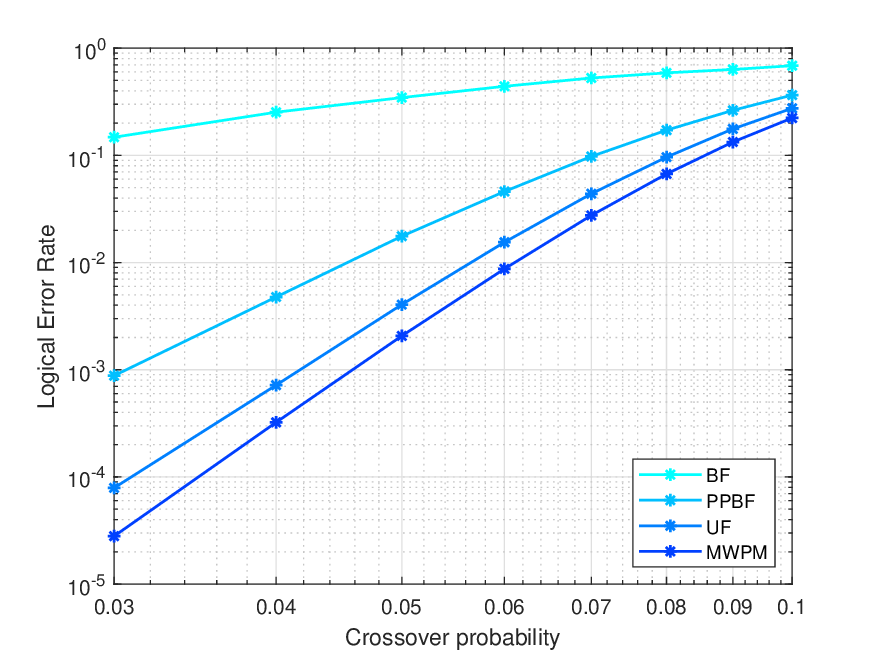}
\caption{Performance comparison between our decoder, traditional BF, MWPM and UF for the distance 13 toric code over the BSC.}
\label{fig:decoder_comparison}
\end{figure}

In this section we present simulation results for toric and rotated surface codes. We perform our simulation assuming a bit-flip channel and assume perfect syndrome measurements, and we compare our decoder with MWPM and UF. We have investigated, through simulations, the best value of $D$ for each rotated and toric code. We found that, as long as $D\geq L$, its value does not impact significantly on the performance of the decoder. The reason for the condition $D\geq L$ is to ensure that all the check and variable nodes have a non-zero proximity vector regardless of the check node of reference; if this is not the case, there might be situations in line 9 of Algorithm \ref{alg:iter-matching} where no target node is detected. For each data point in the plotted curves, the simulation was run until either 100 logical errors were obtained or $10^5$ error vectors were processed.
In Fig.~\ref{fig:rotated} we plot simulation results of our decoder on rotated planar codes over the BSC. The threshold can be seen to occur around $7\%$.
In Fig. \ref{fig:toric} we plot simulation results of our decoder on toric codes over the bit-flip channel. For comparison, the threshold of MWPM on toric code is 10.3\% \cite{dennis_topological_2002}, while that of UF is 9.9\% \cite{delfosse_almost-linear_2021}. As can be seen from the figure, the threshold is around $7.5\%$; to obtain the threshold for the depolarizing channel, it is sufficient to multiply the threshold value by 3/2 \cite{mackay_sparse-graph_2004}. 
In Figs. \ref{fig:comp_rotated} and  \ref{fig:decoder_comparison} we highlight the comparison between PPBF, traditional BF and MWPM for the rotated and toric codes with $L=13$ (for the latter, we also add the performance of the UF decoder for comparison). For the traditional BF, in each iteration we flip all of the bits involved in two unsatisfied checks, and we run it for a maximum of 100 iterations. As expected, MWPM, having higher complexity, achieves the best performance, both in terms of threshold and waterfall; the UF presents slightly worse performance than MWPM, although it has significantly lower complexity. Our decoder exhibits a different slope compared to the other decoders, meaning that it does not take fully advantage of the distance of the code. Nevertheless, it still achieves good performance, comparable to that of MWPM and UF; we also need to stress that our decoder is a hard-decision decoder, while MWPM and UF both utilize soft information. Comparing to the classical BF decoder, the PPBF shows a significantly lower error rate.

\section{Conclusion}
We have presented a new decoder for surface and toric codes which is able to achieve a very good decoding performance with reasonably low decoding complexity, which is particularly suitable for efficient hardware implementation compared to state-of-art decoders. First, we have defined a novel heuristic called the proximity influence, which assigns weights to variable and check nodes in the neighborhood of an unsatisfied check; we also showed that the proximity influence of a check node $c_j$ can be efficiently obtained by appropriately permuting the influence of an arbitrary node $c_1$ which is computed offline and stored in memory. The total contribution of the influences of all of the unsatisfied checks is called the proximity vector, and we exploit it as a metric for flipping bits. To deal with error degeneracy and low variable node degrees of toric and surface codes, we designed the iterative matching procedure, which is employed after a round of serial BF. Future work could include improving the performance of the decoder using decoding diversity, \textit{i.e.}, running several rounds of PPBF, each one using a different variation of the proximity metric, and choosing as error estimate the one with least weight. Also, methods to incorporate soft information on the qubits or the syndrome into the proposed decoding algorithm could lead to performance improvement. Adapting the decoder to work on more general QLDPC codes could also could also be an interesting direction for future research.

\ifCLASSOPTIONcaptionsoff
  \newpage
\fi



\bibliographystyle{IEEEtran}
\input{bibtex/bib/referencesNEW.tex}

%

\end{document}

%% file: tikz_figures/lattice_numbering.tikz
\begin{tikzpicture}[scale=0.75]
    \tikzstyle{var} = [circle, fill=blue!55, draw=black]
        \tikzstyle{var2} = [circle, fill=white, draw=black]

\tikzstyle{varERR} = [circle, fill=red!55, draw=black]
\tikzstyle{checkUNSATI} = [ fill=black]
\tikzstyle{checkSATI} = [ fill=white, draw=black]

\node[checkSATI, label=above:\color{red}{\scriptsize$0$}] (c1) at (2.1,0.7){};
\node[var2, label=above:\scriptsize$0$] (v1) at (0,0){};
\node[var2, label=above:\scriptsize$1$] (v2) at (1.4,0){};
\node[var2, label=above:\scriptsize$2$] (v3) at (2.8,0){};

\node[checkSATI, label=above:\color{red}{\scriptsize$1$}] (c2) at (0.7,-0.7){};

\node[var2, label=above:\scriptsize$3$] (v4) at (0,-1.4){};
\node[var2, label=above:\scriptsize$4$] (v5) at (1.4,-1.4){};
\node[var2, label=above:\scriptsize$5$] (v6) at (2.8,-1.4){};

\node[checkSATI, label=above:\color{red}{\scriptsize$2$}] (c3) at (2.1,-2.1){};

\node[var2, label=above:\scriptsize$6$] (v7) at (0,-2.8){};
\node[var2, label=above:\scriptsize$7$] (v8) at (1.4,-2.8){};
\node[var2, label=above:\scriptsize$8$] (v9) at (2.8,-2.8){};

\node[checkSATI, label=above:\color{red}{\scriptsize$3$}] (c4) at (0.7,-3.5){};

\draw (c1)--(v2);
\draw (c1)--(v3);

\draw (c2)--(v1);
\draw (c2)--(v2);
\draw (c2)--(v4);
\draw (c2)--(v5);

\draw (c3)--(v5);
\draw (c3)--(v6);
\draw (c3)--(v8);
\draw (c3)--(v9);

\draw (c4)--(v7);
\draw (c4)--(v8);

\end{tikzpicture}   

%% file: Algorithms/alg_shift_influence.tex
\begin{algorithmic}[1]
        \State $\mathbf{k}_v, \mathbf{k}_c \gets [1,L_2^2]$
        \State Compute $\sigma_x, \sigma_y$ using (\ref{eq:shift})
        \State Compute $\mathbf{k}'_v, \mathbf{k}'_c$ using (\ref{eq:coord_shift})
        \State $\boldsymbol{\nu}_{\phi_v^{-1}(\mathbf{k}'_v)}(c_j) \gets \boldsymbol{\nu}_{\mathbf{k}_v}(c_{\phi_c(1)}) $
        \State $\boldsymbol{\gamma}_{\phi_c^{-1}(\mathbf{k}'_c)}(c_j) \gets \boldsymbol{\gamma}_{\mathbf{k}_c}(c_{\phi_c(1)}) $
        \State \Return 
  \end{algorithmic}

%% file: tikz_figures/rotated_influece.tikz
\begin{tikzpicture}[scale=1.5]
    \tikzstyle{var} = [circle, fill=white, draw=black, very thick]
        \tikzstyle{var2} = [circle, fill=white, draw=black!55]
    \tikzstyle{varinft} = [circle, fill=grey!55, thick]
     \tikzstyle{varinf} = [circle, fill=grey!55]

\tikzstyle{varERR} = [circle, fill=red!55, draw=black]
\tikzstyle{checkUNSATI} = [ fill=white, draw=black, very thick]
\tikzstyle{checkSATI} = [ fill=white, draw=black!55]

\tikzstyle{checkPIVOT} = [ fill=red, draw=black, thick]

    \tikzstyle{var3} = [circle, fill=orange!35, draw=black]
    \tikzstyle{var4} = [circle, fill=yellow!35, draw=black]

\tikzstyle{varERR} = [circle, fill=red!55, draw=black]
\tikzstyle{checkSATI} = [ fill=white, draw=black]
\tikzstyle{checkSTART} = [ fill=red, draw=black]
\tikzstyle{checkSATI2} = [ fill=white, draw=black!55]
\tikzstyle{checkTARGET} = [ fill=red!35, draw=black]

\node[checkSATI2, label=above:\color{red}{\scriptsize$0$}] (c0) at (0,0){};
\node[var2, label=above:\scriptsize$0$] (v0) at (0.5,0){};
\node[checkSATI2, label=above:\color{red}{\scriptsize$1$}] (c1) at (1,0){};
\node[var2, label=above:\scriptsize$1$] (v1) at (1.5,0){};
\node[checkSATI2, label=above:\color{red}{\scriptsize$2$},fill=gray!55] (c2) at (2,0){};
\node[var2, label=above:\scriptsize$2$] (v2) at (2.5,0){};
\node[checkSATI2, label=above:\color{red}{\scriptsize$3$}] (c3) at (3,0){};
\node[var2, label=above:\scriptsize$3$] (v3) at (3.5,0){};
\node[checkSATI2, label=above:\color{red}{\scriptsize$4$}] (c4) at (4,0){};
\node[var2, label=above:\scriptsize$4$] (v4) at (4.5,0){};
\node[checkSATI2, label=above:\color{red}{\scriptsize$5$}] (c5) at (5,0){};

\node[var2, label=left:\scriptsize$5$] (v5) at (0,-0.5){};
\node[var2, label=left:\scriptsize$6$] (v6) at (1,-0.5){};
\node[var2, label=left:\scriptsize$7$,fill=gray!55] (v7) at (2,-0.5){};
\node[var2, label=left:\scriptsize$8$] (v8) at (3,-0.5){};
\node[var2, label=left:\scriptsize$9$] (v9) at (4,-0.5){};
\node[var2, label=right:\scriptsize$10$] (v10) at (5,-0.5){};

\node[checkSATI2, label=left:\color{red}{\scriptsize$6$}] (c6) at (0,-1){};
\node[var2, label=below:\scriptsize$11$] (v11) at (0.5,-1){};
\node[checkSATI2, label=70:\color{red}{\scriptsize$7$},fill=gray!55] (c7) at (1,-1){};
\node[var2, label=below:\scriptsize$12$,fill=gray!55] (v12) at (1.5,-1){};
\node[checkUNSATI, label=70:\color{red}{\scriptsize$\mathbf{8(0)}$},fill=red] (c8) at (2,-1){};
\node[var, label=below:\scriptsize$\mathbf{13(2)}$,fill=gray!55] (v13) at (2.5,-1){};
\node[checkSATI2, label=70:\color{red}{\scriptsize$9$},fill=gray!55] (c9) at (3,-1){};
\node[var2, label=below:\scriptsize$14$] (v14) at (3.5,-1){};
\node[checkSATI2, label=70:\color{red}{\scriptsize$10$}] (c10) at (4,-1){};
\node[var2, label=below:\scriptsize$15$] (v15) at (4.5,-1){};
\node[checkSATI2, label=right:\color{red}{\scriptsize$11$}] (c11) at (5,-1){};

\node[var2, label=left:\scriptsize$16$] (v16) at (0,-1.5){};
\node[var2, label=left:\scriptsize$17$] (v17) at (1,-1.5){};
\node[var, label=left:\scriptsize$\mathbf{18(1)}$,fill=gray!55] (v18) at (2,-1.5){};
\node[var, label=left:\scriptsize$\mathbf{19(5)}$] (v19) at (3,-1.5){};
\node[var2, label=left:\scriptsize$20$] (v20) at (4,-1.5){};
\node[var2, label=right:\scriptsize$21$] (v21) at (5,-1.5){};

\node[checkSATI2, label=left:\color{red}{\scriptsize$12$}] (c12) at (0,-2){};
\node[var2, label=below:\scriptsize$22$] (v22) at (0.5,-2){};
\node[checkSATI2, label=70:\color{red}{\scriptsize$13$}] (c13) at (1,-2){};
\node[var, label=below:\scriptsize$\mathbf{23(0)}$] (v23) at (1.5,-2){};
\node[checkUNSATI, label=70:\color{red}{\scriptsize$\mathbf{14(1)}$},fill=gray!55] (c14) at (2,-2){};
\node[var, label=below:\scriptsize$\mathbf{24(4)}$] (v24) at (2.5,-2){};
\node[checkUNSATI, label=70:\color{red}{\scriptsize$\mathbf{15(2)}$}] (c15) at (3,-2){};
\node[var, label=below:\scriptsize$\mathbf{25(8)}$] (v25) at (3.5,-2){};
\node[checkSATI2, label=70:\color{red}{\scriptsize$16$}] (c16) at (4,-2){};
\node[var2, label=below:\scriptsize$26$] (v26) at (4.5,-2){};
\node[checkSATI2, label=right:\color{red}{\scriptsize$17$}] (c17) at (5,-2){};

\node[var2, label=left:\scriptsize$27$] (v27) at (0,-2.5){};
\node[var2, label=left:\scriptsize$28$] (v28) at (1,-2.5){};
\node[var, label=left:\scriptsize$\mathbf{29(3)}$] (v29) at (2,-2.5){};
\node[var, label=left:\scriptsize$\mathbf{30(7)}$] (v30) at (3,-2.5){};
\node[var2, label=left:\scriptsize$31$] (v31) at (4,-2.5){};
\node[var2, label=right:\scriptsize$32$] (v32) at (5,-2.5){};

\node[checkSATI2, label=left:\color{red}{\scriptsize$18$}] (c18) at (0,-3){};
\node[var2, label=below:\scriptsize$33$] (v33) at (0.5,-3){};
\node[checkSATI2, label=70:\color{red}{\scriptsize$19$}] (c19) at (1,-3){};
\node[var2, label=below:\scriptsize$34$] (v34) at (1.5,-3){};
\node[checkSATI2, label=70:\color{red}{\scriptsize$20$}] (c20) at (2,-3){};
\node[var, label=below:\scriptsize$\mathbf{35(6)}$] (v35) at (2.5,-3){};
\node[checkUNSATI, label=70:\color{red}{\scriptsize$\mathbf{21(3)}$}] (c21) at (3,-3){};
\node[var2, label=below:\scriptsize$36$] (v36) at (3.5,-3){};
\node[checkSATI2, label=70:\color{red}{\scriptsize$22$}] (c22) at (4,-3){};
\node[var2, label=below:\scriptsize$37$] (v37) at (4.5,-3){};
\node[checkSATI2, label=right:\color{red}{\scriptsize$23$}] (c23) at (5,-3){};

\node[var2, label=left:\scriptsize$38$] (v38) at (0,-3.5){};
\node[var2, label=left:\scriptsize$39$] (v39) at (1,-3.5){};
\node[var2, label=left:\scriptsize$40$] (v40) at (2,-3.5){};
\node[var2, label=left:\scriptsize$41$] (v41) at (3,-3.5){};
\node[var2, label=left:\scriptsize$42$] (v42) at (4,-3.5){};
\node[var2, label=right:\scriptsize$43$] (v43) at (5,-3.5){};

\node[checkSATI2, label=left:\color{red}{\scriptsize$24$}] (c24) at (0,-4){};
\node[var2, label=below:\scriptsize$44$] (v44) at (0.5,-4){};
\node[checkSATI2, label=70:\color{red}{\scriptsize$25$}] (c25) at (1,-4){};
\node[var2, label=below:\scriptsize$45$] (v45) at (1.5,-4){};
\node[checkSATI2, label=70:\color{red}{\scriptsize$26$}] (c26) at (2,-4){};
\node[var2, label=below:\scriptsize$46$] (v46) at (2.5,-4){};
\node[checkSATI2, label=70:\color{red}{\scriptsize$27$}] (c27) at (3,-4){};
\node[var2, label=below:\scriptsize$47$] (v47) at (3.5,-4){};
\node[checkSATI2, label=70:\color{red}{\scriptsize$28$}] (c28) at (4,-4){};
\node[var2, label=below:\scriptsize$48$] (v48) at (4.5,-4){};
\node[checkSATI2, label=right:\color{red}{\scriptsize$29$}] (c29) at (5,-4){};



\draw[black!35] (c0)--(v0);
\draw[black!35] (c1)--(v0);
\draw[black!35] (c1)--(v1);
\draw[black!35] (c2)--(v1);
\draw[black!35] (c2)--(v2);
\draw[black!35] (c3)--(v2);
\draw[black!35] (c3)--(v3);
\draw[black!35] (c4)--(v3);
\draw[black!35] (c4)--(v4);
\draw[black!35] (c5)--(v4);

\draw[black!35] (c0)--(v5);
\draw[black!35] (c6)--(v5);
\draw[black!35] (c1)--(v6);
\draw[black!35] (c7)--(v6);
\draw[black!35] (c2)--(v7);
\draw[black!35] (c8)--(v7);
\draw[black!35] (c3)--(v8);
\draw[black!35] (c9)--(v8);
\draw[black!35] (c4)--(v9);
\draw[black!35] (c10)--(v9);
\draw[black!35] (c11)--(v10);
\draw[black!35] (c5)--(v10);

\draw[black!35] (c6)--(v11);
\draw[black!35] (c7)--(v11);
\draw[black!35] (c7)--(v12);
\draw[black!35] (c8)--(v12);
\draw[very thick] (c8)--(v13);
\draw[black!35] (c9)--(v13);
\draw[black!35] (c9)--(v14);
\draw[black!35] (c10)--(v14);
\draw[black!35] (c10)--(v15);
\draw[black!35] (c11)--(v15);

\draw[black!35] (c6)--(v16);
\draw[black!35] (c12)--(v16);
\draw[black!35] (c7)--(v17);
\draw[black!35] (c13)--(v17);
\draw[very thick] (c8)--(v18);
\draw[very thick] (c14)--(v18);
\draw[black!35] (c9)--(v19);
\draw[very thick] (c15)--(v19);
\draw[black!35] (c10)--(v20);
\draw[black!35] (c16)--(v20);
\draw[black!35] (c11)--(v21);
\draw[black!35] (c17)--(v21);

\draw[black!35] (c12)--(v22);
\draw[black!35] (c13)--(v22);
\draw[black!35] (c13)--(v23);
\draw[very thick] (c14)--(v23);
\draw[very thick] (c14)--(v24);
\draw[very thick] (c15)--(v24);
\draw[very thick] (c15)--(v25);
\draw[black!35] (c16)--(v25);
\draw[black!35] (c16)--(v26);
\draw[black!35] (c17)--(v26);

\draw[black!35] (c12)--(v27);
\draw[black!35] (c18)--(v27);
\draw[black!35] (c13)--(v28);
\draw[black!35] (c19)--(v28);
\draw[very thick] (c14)--(v29);
\draw[black!35] (c20)--(v29);
\draw[very thick] (c15)--(v30);
\draw[very thick] (c21)--(v30);
\draw[black!35] (c16)--(v31);
\draw[black!35] (c22)--(v31);
\draw[black!35] (c17)--(v32);
\draw[black!35] (c23)--(v32);

\draw[black!35] (c18)--(v33);
\draw[black!35] (c19)--(v33);
\draw[black!35] (c19)--(v34);
\draw[black!35] (c20)--(v34);
\draw[black!35] (c20)--(v35);
\draw[very thick] (c21)--(v35);
\draw[black!35] (c21)--(v36);
\draw[black!35] (c22)--(v36);
\draw[black!35] (c22)--(v37);
\draw[black!35] (c23)--(v37);

\draw[black!35] (c18)--(v38);
\draw[black!35] (c24)--(v38);
\draw[black!35] (c19)--(v39);
\draw[black!35] (c25)--(v39);
\draw[black!35] (c20)--(v40);
\draw[black!35] (c26)--(v40);
\draw[black!35] (c21)--(v41);
\draw[black!35] (c27)--(v41);
\draw[black!35] (c22)--(v42);
\draw[black!35] (c28)--(v42);
\draw[black!35] (c23)--(v43);
\draw[black!35] (c29)--(v43);

\draw[black!35] (c24)--(v44);
\draw[black!35] (c25)--(v44);
\draw[black!35] (c25)--(v45);
\draw[black!35] (c26)--(v45);
\draw[black!35] (c26)--(v46);
\draw[black!35] (c27)--(v46);
\draw[black!35] (c27)--(v47);
\draw[black!35] (c28)--(v47);
\draw[black!35] (c28)--(v48);
\draw[black!35] (c29)--(v48);



\end{tikzpicture}   

%% file: tikz_figures/rotated_influence_2.tikz
\begin{tikzpicture}[scale=1.5]
    \tikzstyle{var} = [circle, fill=white, draw=black, very thick]
        \tikzstyle{var2} = [circle, fill=white, draw=black!55]
    \tikzstyle{varinft} = [circle, fill=grey!55, thick]
     \tikzstyle{varinf} = [circle, fill=grey!55]

\tikzstyle{varERR} = [circle, fill=red!55, draw=black]
\tikzstyle{checkUNSATI} = [ fill=white, draw=black, very thick]
\tikzstyle{checkSATI} = [ fill=white, draw=black!55]

\tikzstyle{checkPIVOT} = [ fill=red, draw=black, thick]

    \tikzstyle{var3} = [circle, fill=orange!35, draw=black]
    \tikzstyle{var4} = [circle, fill=yellow!35, draw=black]

\tikzstyle{varERR} = [circle, fill=red!55, draw=black]
\tikzstyle{checkSATI} = [ fill=white, draw=black]
\tikzstyle{checkSTART} = [ fill=red, draw=black]
\tikzstyle{checkSATI2} = [ fill=white, draw=black!55]
\tikzstyle{checkTARGET} = [ fill=red!35, draw=black]

\node[checkSATI2, label=above:\color{red}{\scriptsize$0$}] (c0) at (0,0){};
\node[var2, label=above:\scriptsize$0$] (v0) at (0.5,0){};
\node[checkSATI2, label=above:\color{red}{\scriptsize$1$}] (c1) at (1,0){};
\node[var2, label=above:\scriptsize$1$] (v1) at (1.5,0){};
\node[checkSATI2, label=above:\color{red}{\scriptsize$2$}] (c2) at (2,0){};
\node[var2, label=above:\scriptsize$2$] (v2) at (2.5,0){};
\node[checkSATI2, label=above:\color{red}{\scriptsize$3$}] (c3) at (3,0){};
\node[var2, label=above:\scriptsize$3$] (v3) at (3.5,0){};
\node[checkSATI2, label=above:\color{red}{\scriptsize$4$}] (c4) at (4,0){};
\node[var2, label=above:\scriptsize$4$] (v4) at (4.5,0){};
\node[checkSATI2, label=above:\color{red}{\scriptsize$5$}] (c5) at (5,0){};

\node[var2, label=left:\scriptsize$5$] (v5) at (0,-0.5){};
\node[var2, label=left:\scriptsize$6$] (v6) at (1,-0.5){};
\node[var2, label=left:\scriptsize$7$,] (v7) at (2,-0.5){};
\node[var2, label=left:\scriptsize$8$] (v8) at (3,-0.5){};
\node[var2, label=left:\scriptsize$9$] (v9) at (4,-0.5){};
\node[var2, label=right:\scriptsize$10$] (v10) at (5,-0.5){};

\node[checkSATI2, label=left:\color{red}{\scriptsize$6$}] (c6) at (0,-1){};
\node[var2, label=below:\scriptsize$11$] (v11) at (0.5,-1){};
\node[checkSATI2, label=70:\color{red}{\scriptsize$7$},] (c7) at (1,-1){};
\node[var2, label=below:\scriptsize$12$,] (v12) at (1.5,-1){};
\node[checkUNSATI, label=70:\color{red}{\scriptsize$\mathbf{8(0)}$}] (c8) at (2,-1){};
\node[var, label=below:\scriptsize$\mathbf{13(2)}$,] (v13) at (2.5,-1){};
\node[checkSATI2, label=70:\color{red}{\scriptsize$9$},fill=gray!55] (c9) at (3,-1){};
\node[var2, label=below:\scriptsize$14$] (v14) at (3.5,-1){};
\node[checkSATI2, label=70:\color{red}{\scriptsize$10$}] (c10) at (4,-1){};
\node[var2, label=below:\scriptsize$15$] (v15) at (4.5,-1){};
\node[checkSATI2, label=right:\color{red}{\scriptsize$11$}] (c11) at (5,-1){};

\node[var2, label=left:\scriptsize$16$] (v16) at (0,-1.5){};
\node[var2, label=left:\scriptsize$17$] (v17) at (1,-1.5){};
\node[var, label=left:\scriptsize$\mathbf{18(1)}$,] (v18) at (2,-1.5){};
\node[var, label=left:\scriptsize$\mathbf{19(5)}$,fill=gray!55] (v19) at (3,-1.5){};
\node[var2, label=left:\scriptsize$20$] (v20) at (4,-1.5){};
\node[var2, label=right:\scriptsize$21$] (v21) at (5,-1.5){};

\node[checkSATI2, label=left:\color{red}{\scriptsize$12$}] (c12) at (0,-2){};
\node[var2, label=below:\scriptsize$22$] (v22) at (0.5,-2){};
\node[checkSATI2, label=70:\color{red}{\scriptsize$13$}] (c13) at (1,-2){};
\node[var, label=below:\scriptsize$\mathbf{23(0)}$] (v23) at (1.5,-2){};
\node[checkUNSATI, label=70:\color{red}{\scriptsize$\mathbf{14(1)}$},fill=gray!55] (c14) at (2,-2){};
\node[var, label=below:\scriptsize$\mathbf{24(4)}$,fill=gray!55] (v24) at (2.5,-2){};
\node[checkUNSATI, label=70:\color{red}{\scriptsize$\mathbf{15(2)}$},fill=red] (c15) at (3,-2){};
\node[var, label=below:\scriptsize$\mathbf{25(8)}$,fill=gray!55] (v25) at (3.5,-2){};
\node[checkSATI2, label=70:\color{red}{\scriptsize$16$},fill=gray!55] (c16) at (4,-2){};
\node[var2, label=below:\scriptsize$26$] (v26) at (4.5,-2){};
\node[checkSATI2, label=right:\color{red}{\scriptsize$17$}] (c17) at (5,-2){};

\node[var2, label=left:\scriptsize$27$] (v27) at (0,-2.5){};
\node[var2, label=left:\scriptsize$28$] (v28) at (1,-2.5){};
\node[var, label=left:\scriptsize$\mathbf{29(3)}$] (v29) at (2,-2.5){};
\node[var, label=left:\scriptsize$\mathbf{30(7)}$,fill=gray!55] (v30) at (3,-2.5){};
\node[var2, label=left:\scriptsize$31$] (v31) at (4,-2.5){};
\node[var2, label=right:\scriptsize$32$] (v32) at (5,-2.5){};

\node[checkSATI2, label=left:\color{red}{\scriptsize$18$}] (c18) at (0,-3){};
\node[var2, label=below:\scriptsize$33$] (v33) at (0.5,-3){};
\node[checkSATI2, label=70:\color{red}{\scriptsize$19$}] (c19) at (1,-3){};
\node[var2, label=below:\scriptsize$34$] (v34) at (1.5,-3){};
\node[checkSATI2, label=70:\color{red}{\scriptsize$20$}] (c20) at (2,-3){};
\node[var, label=below:\scriptsize$\mathbf{35(6)}$] (v35) at (2.5,-3){};
\node[checkUNSATI, label=70:\color{red}{\scriptsize$\mathbf{21(3)}$},fill=gray!55] (c21) at (3,-3){};
\node[var2, label=below:\scriptsize$36$] (v36) at (3.5,-3){};
\node[checkSATI2, label=70:\color{red}{\scriptsize$22$}] (c22) at (4,-3){};
\node[var2, label=below:\scriptsize$37$] (v37) at (4.5,-3){};
\node[checkSATI2, label=right:\color{red}{\scriptsize$23$}] (c23) at (5,-3){};

\node[var2, label=left:\scriptsize$38$] (v38) at (0,-3.5){};
\node[var2, label=left:\scriptsize$39$] (v39) at (1,-3.5){};
\node[var2, label=left:\scriptsize$40$] (v40) at (2,-3.5){};
\node[var2, label=left:\scriptsize$41$] (v41) at (3,-3.5){};
\node[var2, label=left:\scriptsize$42$] (v42) at (4,-3.5){};
\node[var2, label=right:\scriptsize$43$] (v43) at (5,-3.5){};

\node[checkSATI2, label=left:\color{red}{\scriptsize$24$}] (c24) at (0,-4){};
\node[var2, label=below:\scriptsize$44$] (v44) at (0.5,-4){};
\node[checkSATI2, label=70:\color{red}{\scriptsize$25$}] (c25) at (1,-4){};
\node[var2, label=below:\scriptsize$45$] (v45) at (1.5,-4){};
\node[checkSATI2, label=70:\color{red}{\scriptsize$26$}] (c26) at (2,-4){};
\node[var2, label=below:\scriptsize$46$] (v46) at (2.5,-4){};
\node[checkSATI2, label=70:\color{red}{\scriptsize$27$}] (c27) at (3,-4){};
\node[var2, label=below:\scriptsize$47$] (v47) at (3.5,-4){};
\node[checkSATI2, label=70:\color{red}{\scriptsize$28$}] (c28) at (4,-4){};
\node[var2, label=below:\scriptsize$48$] (v48) at (4.5,-4){};
\node[checkSATI2, label=right:\color{red}{\scriptsize$29$}] (c29) at (5,-4){};



\draw[black!35] (c0)--(v0);
\draw[black!35] (c1)--(v0);
\draw[black!35] (c1)--(v1);
\draw[black!35] (c2)--(v1);
\draw[black!35] (c2)--(v2);
\draw[black!35] (c3)--(v2);
\draw[black!35] (c3)--(v3);
\draw[black!35] (c4)--(v3);
\draw[black!35] (c4)--(v4);
\draw[black!35] (c5)--(v4);

\draw[black!35] (c0)--(v5);
\draw[black!35] (c6)--(v5);
\draw[black!35] (c1)--(v6);
\draw[black!35] (c7)--(v6);
\draw[black!35] (c2)--(v7);
\draw[black!35] (c8)--(v7);
\draw[black!35] (c3)--(v8);
\draw[black!35] (c9)--(v8);
\draw[black!35] (c4)--(v9);
\draw[black!35] (c10)--(v9);
\draw[black!35] (c11)--(v10);
\draw[black!35] (c5)--(v10);

\draw[black!35] (c6)--(v11);
\draw[black!35] (c7)--(v11);
\draw[black!35] (c7)--(v12);
\draw[black!35] (c8)--(v12);
\draw[very thick] (c8)--(v13);
\draw[black!35] (c9)--(v13);
\draw[black!35] (c9)--(v14);
\draw[black!35] (c10)--(v14);
\draw[black!35] (c10)--(v15);
\draw[black!35] (c11)--(v15);

\draw[black!35] (c6)--(v16);
\draw[black!35] (c12)--(v16);
\draw[black!35] (c7)--(v17);
\draw[black!35] (c13)--(v17);
\draw[very thick] (c8)--(v18);
\draw[very thick] (c14)--(v18);
\draw[black!35] (c9)--(v19);
\draw[very thick] (c15)--(v19);
\draw[black!35] (c10)--(v20);
\draw[black!35] (c16)--(v20);
\draw[black!35] (c11)--(v21);
\draw[black!35] (c17)--(v21);

\draw[black!35] (c12)--(v22);
\draw[black!35] (c13)--(v22);
\draw[black!35] (c13)--(v23);
\draw[very thick] (c14)--(v23);
\draw[very thick] (c14)--(v24);
\draw[very thick] (c15)--(v24);
\draw[very thick] (c15)--(v25);
\draw[black!35] (c16)--(v25);
\draw[black!35] (c16)--(v26);
\draw[black!35] (c17)--(v26);

\draw[black!35] (c12)--(v27);
\draw[black!35] (c18)--(v27);
\draw[black!35] (c13)--(v28);
\draw[black!35] (c19)--(v28);
\draw[very thick] (c14)--(v29);
\draw[black!35] (c20)--(v29);
\draw[very thick] (c15)--(v30);
\draw[very thick] (c21)--(v30);
\draw[black!35] (c16)--(v31);
\draw[black!35] (c22)--(v31);
\draw[black!35] (c17)--(v32);
\draw[black!35] (c23)--(v32);

\draw[black!35] (c18)--(v33);
\draw[black!35] (c19)--(v33);
\draw[black!35] (c19)--(v34);
\draw[black!35] (c20)--(v34);
\draw[black!35] (c20)--(v35);
\draw[very thick] (c21)--(v35);
\draw[black!35] (c21)--(v36);
\draw[black!35] (c22)--(v36);
\draw[black!35] (c22)--(v37);
\draw[black!35] (c23)--(v37);

\draw[black!35] (c18)--(v38);
\draw[black!35] (c24)--(v38);
\draw[black!35] (c19)--(v39);
\draw[black!35] (c25)--(v39);
\draw[black!35] (c20)--(v40);
\draw[black!35] (c26)--(v40);
\draw[black!35] (c21)--(v41);
\draw[black!35] (c27)--(v41);
\draw[black!35] (c22)--(v42);
\draw[black!35] (c28)--(v42);
\draw[black!35] (c23)--(v43);
\draw[black!35] (c29)--(v43);

\draw[black!35] (c24)--(v44);
\draw[black!35] (c25)--(v44);
\draw[black!35] (c25)--(v45);
\draw[black!35] (c26)--(v45);
\draw[black!35] (c26)--(v46);
\draw[black!35] (c27)--(v46);
\draw[black!35] (c27)--(v47);
\draw[black!35] (c28)--(v47);
\draw[black!35] (c28)--(v48);
\draw[black!35] (c29)--(v48);



\end{tikzpicture}   

%% file: Algorithms/alg_compute_proximity.tex
 \begin{algorithmic}[1]
        \State $\mathbf{k} \gets k\ |\ s_k=1$
        \State $\boldsymbol{\nu}, \boldsymbol{\gamma} \gets \mathbf{0}$
        \For{$k \in \mathbf{k}$}
            \State $\boldsymbol{\gamma}(c_k), \boldsymbol{\nu}(c_k) \gets \mathtt{Shift\_influence}(k)$
            \State $\boldsymbol{\gamma} \gets \boldsymbol{\gamma} +\boldsymbol{\gamma}(c_k) $
            \State $\boldsymbol{\nu} \gets \boldsymbol{\nu} + \boldsymbol{\nu}(c_k)$
        \EndFor
        \State \Return
  \end{algorithmic}

%% file: Algorithms/alg_shift_and_remove.tex
 \begin{algorithmic}[1]
        \State $\mathbf{j} \gets j\ |\ s'_j=1$
        \State $\boldsymbol{\nu}' \gets \boldsymbol{\nu}$
        \State $\boldsymbol{\gamma}' \gets \boldsymbol{\gamma}$
        \For{$k \in \mathbf{j}$}
            \State $\boldsymbol{\gamma}(c_k), \boldsymbol{\nu}(c_k) \gets \mathtt{Shift\_influence}(k)$
            \State $\boldsymbol{\nu}' \gets \boldsymbol{\nu}' - \boldsymbol{\nu}(c_k)$
            \State $\boldsymbol{\gamma}' \gets \boldsymbol{\gamma}' - \boldsymbol{\gamma}(c_k)$
        \EndFor
        \State \Return
  \end{algorithmic}

%% file: Algorithms/alg_bit_matching.tex
    \begin{algorithmic}[1]
      \State $\boldsymbol{\nu}, \boldsymbol{\gamma} \gets \mathtt{Compute\_proximity\_metric}(\mathbf{s})$
        \State $\hat{\mathbf{e}},\hat{\mathbf{s}},\boldsymbol{\nu}' \gets \mathtt{Preliminary\_BF}
        (\mathbf{s},\mathbf{H},\boldsymbol{\nu})$  
        \State $\hat{\mathbf{e}} \gets \mathtt{Iterative\_matching}(\hat{\mathbf{s}},\mathbf{H},\boldsymbol{\gamma})$
        \State \Return
    \end{algorithmic}

%% file: Algorithms/alg_preliminary_bf.tex
  \begin{algorithmic}[1]
      \State $\hat{\mathbf{e}} \gets \mathbf{0}$
      \State $\hat{\mathbf{s}} \gets \mathbf{s}$
      \State $\mathbf{u} \gets \mathbf{s} \cdot \mathbf{H}$ 
      \label{alg1:counters}
      \State $\boldsymbol{\nu}' \gets \boldsymbol{\nu}$
      \State $\mathcal{S} \gets \{i \in [1,n]: u_i =2\}$
      \While{$\mathcal{S} \ne \emptyset$} \label{alg1:stopping}
        \State $j \gets  \underset{i \in \mathcal{S}}{\mathrm{arg\ min}}\ \nu_i'$ \label{alg1:minproxy}
        \State $\hat{e}_j \gets \hat{e}_j \oplus 1$ \label{alg1:flip}
        \State $\mathbf{s}' \gets \hat{\mathbf{e}}\cdot \mathbf{H}^{T} \mod{2}$
        \State $\boldsymbol{\nu}' \gets \mathtt{Shift\_and\_remove(\boldsymbol{\nu}',\cdot ,\mathbf{s}')}$ \label{alg1:shiftremove}
        \State $\hat{\mathbf{s}} \gets \mathbf{s} \oplus \mathbf{s}'$ \label{alg1:residual}
        \State $\mathbf{u} \gets \hat{\mathbf{s}} \cdot \mathbf{H}$
    \EndWhile
    \State \Return $\hat{\mathbf{e}}$, $\hat{\mathbf{s}}$
  \end{algorithmic}

%% file: tikz_figures/matching_rotated_1.tikz
\begin{tikzpicture}[scale=0.47]
    \tikzstyle{var} = [circle, fill=white, draw=black]
        \tikzstyle{var2} = [circle, fill=white, draw=black]

\tikzstyle{varERR} = [circle, fill=red!55, draw=black]
\tikzstyle{checkUNSATI} = [ fill=white, draw=black,   ]
\tikzstyle{checkSATI} = [ fill=white, draw=black]
\tikzstyle{checkDUM} = [ fill=black, draw=black]

\tikzstyle{checkPIVOT} = [ fill=red, draw=black,   ]

\node[checkSATI, label=above:\color{red}{\scriptsize$0$}] (c11) at (-2,1){};
\node[checkSATI, label=above:\color{red}{\scriptsize$1$}] (c12) at (2,1){};

\node[var2,  label=above:\scriptsize$0$] (v11) at (-5,0){};
\node[var2, label=above:\scriptsize$1$] (v12) at (-3,0){};
\node[var2, label=above:\scriptsize$2$] (v13) at (-1,0){};
\node[var2, label=above:\scriptsize$3$] (v14) at (1,0){};
\node[var2, label=above:\scriptsize$4$] (v15) at (3,0){};

\node[checkSATI,label=above:\color{red}{\scriptsize$2$}] (c21) at (-4,-1){};
\node[checkUNSATI,label=above:\color{red}{\scriptsize$3$}] (c22) at (0,-1){};

\node[var2, label=above:\scriptsize$5$] (v21) at (-5,-2){};
\node[var,  fill=yellow!35, label=above:\scriptsize$6$] (v22) at (-3,-2){};
\node[var,   fill=yellow!35, label=above:\scriptsize$7$] (v23) at (-1,-2){};
\node[var,   fill=yellow!35,label=above:\scriptsize$8$] (v24) at (1,-2){};
\node[var2,  fill=yellow!35, label=above:\scriptsize$9$] (v25) at (3,-2){};

\node[checkUNSATI,fill=gray!55, label=above:\color{red}{\scriptsize$4$}] (c31) at (-2,-3){};
\node[checkSATI, fill=gray!55, label=above:\color{red}{\scriptsize$5$}] (c32) at (2,-3){};

\node[var2, label=above:\scriptsize$10$] (v31) at (-5,-4){};
\node[var,  fill=yellow!35, label=above:\scriptsize$11$] (v32) at (-3,-4){};
\node[var,fill=orange!55, label=above:\scriptsize$12$] (v33) at (-1,-4){};
\node[var,fill=orange!55, label=above:\scriptsize$13$] (v34) at (1,-4){};
\node[var2,  fill=yellow!35, label=above:\scriptsize$14$] (v35) at (3,-4){};

\node[checkSATI, label=above:\color{red}{\scriptsize$6$}] (c41) at (-4,-5){};
\node[checkPIVOT, label=above:\color{red}{\scriptsize$7$}] (c42) at (0,-5){};

\node[var2, label=above:\scriptsize$15$] (v41) at (-5,-6){};
\node[var,  fill=yellow!35, label=above:\scriptsize$16$] (v42) at (-3,-6){};
\node[var,fill=orange!55, label=above:\scriptsize$17$] (v43) at (-1,-6){};
\node[var,fill=orange!55, label=above:\scriptsize$18$] (v44) at (1,-6){};
\node[var2,  fill=yellow!35, label=above:\scriptsize$19$] (v45) at (3,-6){};

\node[checkUNSATI,fill=gray!55, label=above:\color{red}{\scriptsize$8$}] (c51) at (-2,-7){};
\node[checkSATI,fill=gray!55, label=above:\color{red}{\scriptsize$9$}] (c52) at (2,-7){};

\node[var2, label=above:\scriptsize$20$] (v51) at (-5,-8){};
\node[var2,  fill=yellow!35,label=above:\scriptsize$21$] (v52) at (-3,-8){};
\node[var2,  fill=yellow!35,label=above:\scriptsize$22$] (v53) at (-1,-8){};
\node[var2,  fill=yellow!35, label=above:\scriptsize$23$] (v54) at (1,-8){};
\node[var2,  fill=yellow!35, label=above:\scriptsize$24$] (v55) at (3,-8){};

\node[checkSATI, label=above:\color{red}{\scriptsize$10$}] (c61) at (-4,-9){};
\node[checkSATI, label=above:\color{red}{\scriptsize$11$}] (c62) at (0,-9){};

\node[checkDUM, label=above:\color{red}] (ca1) at (4,-1){};
\node[checkDUM, label=above:\color{red}] (ca2) at (4,-5){};
\node[checkDUM, label=above:\color{red}] (ca3) at (-6,-3){};
\node[checkDUM, label=above:\color{red}] (ca4) at (-6,-7){};

\draw (ca1)--(v15) [dashed];
\draw (ca1)--(v25) [dashed];

\draw (ca2)--(v35) [dashed];
\draw (ca2)--(v45) [dashed];

\draw (ca3)--(v21) [dashed];
\draw (ca3)--(v31) [dashed];

\draw (ca4)--(v41) [dashed];
\draw (ca4)--(v51) [dashed];

\draw (c11)--(v12);
\draw (c11)--(v13);
\draw (c12)--(v14);
\draw (c12)--(v15);

\draw (c21)--(v11);
\draw  (c21)--(v12);
\draw  (c21)--(v21);
\draw  (c21)--(v22);

\draw  (c22)--(v13);
\draw  (c22)--(v14);
\draw[  ] (c22)--(v23);
\draw[  ] (c22)--(v24);

\draw[  ] (c31)--(v22);
\draw[  ] (c31)--(v23);
\draw[  ] (c31)--(v32);
\draw[  ] (c31)--(v33);

\draw  (c32)--(v24);
\draw  (c32)--(v25);
\draw  (c32)--(v34);
\draw  (c32)--(v35);

\draw  (c41)--(v31);
\draw  (c41)--(v32);
\draw  (c41)--(v41);
\draw  (c41)--(v42);

\draw[  ] (c42)--(v33);
\draw[  ] (c42)--(v34);
\draw[  ] (c42)--(v43);
\draw[  ] (c42)--(v44);

\draw[  ] (c51)--(v42);
\draw[  ] (c51)--(v43);
\draw  (c51)--(v52);
\draw  (c51)--(v53);

\draw  (c52)--(v44);
\draw  (c52)--(v45);
\draw  (c52)--(v54);
\draw  (c52)--(v55);

\draw  (c61)--(v51);
\draw  (c61)--(v52);

\draw  (c62)--(v53);
\draw  (c62)--(v54);

\end{tikzpicture}   

%% file: tikz_figures/matching_rotated_2.tikz
\begin{tikzpicture}[scale=0.47]
    \tikzstyle{var} = [circle, fill=white, draw=black]
        \tikzstyle{var2} = [circle, fill=white, draw=black]

\tikzstyle{varERR} = [circle, fill=red!55, draw=black]
\tikzstyle{checkUNSATI} = [ fill=white, draw=black,   ]
\tikzstyle{checkSATI} = [ fill=white, draw=black]
\tikzstyle{checkDUM} = [ fill=black, draw=black]

\tikzstyle{checkPIVOT} = [ fill=red, draw=black,   ]

\node[checkSATI,fill=gray!55, label=above:\color{red}{\scriptsize$0$}] (c11) at (-2,1){};
\node[checkSATI, fill=gray!55, label=above:\color{red}{\scriptsize$1$}] (c12) at (2,1){};

\node[var2,  label=above:\scriptsize$0$] (v11) at (-5,0){};
\node[var2,  fill=yellow!35, label=above:\scriptsize$1$] (v12) at (-3,0){};
\node[var2,fill=orange!55, label=above:\scriptsize$2$] (v13) at (-1,0){};
\node[var2,fill=orange!55, label=above:\scriptsize$3$] (v14) at (1,0){};
\node[var2,  fill=yellow!35, label=above:\scriptsize$4$] (v15) at (3,0){};

\node[checkSATI,label=above:\color{red}{\scriptsize$2$}] (c21) at (-4,-1){};
\node[checkPIVOT,label=above:\color{red}{\scriptsize$3$}] (c22) at (0,-1){};

\node[var2, label=above:\scriptsize$5$] (v21) at (-5,-2){};
\node[var,   fill=yellow!35, label=above:\scriptsize$6$] (v22) at (-3,-2){};
\node[var, fill=orange!55, label=above:\scriptsize$7$] (v23) at (-1,-2){};
\node[var, fill=orange!55,label=above:\scriptsize$8$] (v24) at (1,-2){};
\node[var2,  fill=yellow!35, label=above:\scriptsize$9$] (v25) at (3,-2){};

\node[checkUNSATI,fill=gray!55, label=above:\color{red}{\scriptsize$4$}] (c31) at (-2,-3){};
\node[checkSATI, fill=gray!55, label=above:\color{red}{\scriptsize$5$}] (c32) at (2,-3){};

\node[var2, label=above:\scriptsize$10$] (v31) at (-5,-4){};
\node[var,  fill=yellow!35, label=above:\scriptsize$11$] (v32) at (-3,-4){};
\node[var,  fill=yellow!35,label=above:\scriptsize$12$] (v33) at (-1,-4){};
\node[var,  fill=yellow!35,label=above:\scriptsize$13$] (v34) at (1,-4){};
\node[var2,  fill=yellow!35, label=above:\scriptsize$14$] (v35) at (3,-4){};

\node[checkSATI, label=above:\color{red}{\scriptsize$6$}] (c41) at (-4,-5){};
\node[checkSATI, label=above:\color{red}{\scriptsize$7$}] (c42) at (0,-5){};

\node[var2, label=above:\scriptsize$15$] (v41) at (-5,-6){};
\node[var,label=above:\scriptsize$16$] (v42) at (-3,-6){};
\node[var,label=above:\scriptsize$17$] (v43) at (-1,-6){};
\node[var,label=above:\scriptsize$18$] (v44) at (1,-6){};
\node[var2,label=above:\scriptsize$19$] (v45) at (3,-6){};

\node[checkUNSATI, label=above:\color{red}{\scriptsize$8$}] (c51) at (-2,-7){};
\node[checkSATI, label=above:\color{red}{\scriptsize$9$}] (c52) at (2,-7){};

\node[var2, label=above:\scriptsize$20$] (v51) at (-5,-8){};
\node[var2,label=above:\scriptsize$21$] (v52) at (-3,-8){};
\node[var2,label=above:\scriptsize$22$] (v53) at (-1,-8){};
\node[var2, label=above:\scriptsize$23$] (v54) at (1,-8){};
\node[var2, label=above:\scriptsize$24$] (v55) at (3,-8){};

\node[checkSATI, label=above:\color{red}{\scriptsize$10$}] (c61) at (-4,-9){};
\node[checkSATI, label=above:\color{red}{\scriptsize$11$}] (c62) at (0,-9){};

\node[checkDUM, label=above:\color{red}] (ca1) at (4,-1){};
\node[checkDUM, label=above:\color{red}] (ca2) at (4,-5){};
\node[checkDUM, label=above:\color{red}] (ca3) at (-6,-3){};
\node[checkDUM, label=above:\color{red}] (ca4) at (-6,-7){};

\draw (ca1)--(v15) [dashed];
\draw (ca1)--(v25) [dashed];

\draw (ca2)--(v35) [dashed];
\draw (ca2)--(v45) [dashed];

\draw (ca3)--(v21) [dashed];
\draw (ca3)--(v31) [dashed];

\draw (ca4)--(v41) [dashed];
\draw (ca4)--(v51) [dashed];

\draw (c11)--(v12);
\draw (c11)--(v13);
\draw (c12)--(v14);
\draw (c12)--(v15);

\draw (c21)--(v11);
\draw  (c21)--(v12);
\draw  (c21)--(v21);
\draw  (c21)--(v22);

\draw  (c22)--(v13);
\draw  (c22)--(v14);
\draw[  ] (c22)--(v23);
\draw[  ] (c22)--(v24);

\draw[  ] (c31)--(v22);
\draw[  ] (c31)--(v23);
\draw[  ] (c31)--(v32);
\draw[  ] (c31)--(v33);

\draw  (c32)--(v24);
\draw  (c32)--(v25);
\draw  (c32)--(v34);
\draw  (c32)--(v35);

\draw  (c41)--(v31);
\draw  (c41)--(v32);
\draw  (c41)--(v41);
\draw  (c41)--(v42);

\draw[  ] (c42)--(v33);
\draw[  ] (c42)--(v34);
\draw[  ] (c42)--(v43);
\draw[  ] (c42)--(v44);

\draw[  ] (c51)--(v42);
\draw[  ] (c51)--(v43);
\draw  (c51)--(v52);
\draw  (c51)--(v53);

\draw  (c52)--(v44);
\draw  (c52)--(v45);
\draw  (c52)--(v54);
\draw  (c52)--(v55);

\draw  (c61)--(v51);
\draw  (c61)--(v52);

\draw  (c62)--(v53);
\draw  (c62)--(v54);

\end{tikzpicture}   

%% file: tikz_figures/matching_rotated_3.tikz
\begin{tikzpicture}[scale=0.47]
    \tikzstyle{var} = [circle, fill=white, draw=black]
        \tikzstyle{var2} = [circle, fill=white, draw=black]

\tikzstyle{varERR} = [circle, fill=red!55, draw=black]
\tikzstyle{checkUNSATI} = [ fill=white, draw=black,   ]
\tikzstyle{checkSATI} = [ fill=white, draw=black]
\tikzstyle{checkDUM} = [ fill=black, draw=black]

\tikzstyle{checkPIVOT} = [ fill=red, draw=black,   ]

\node[checkSATI,fill=gray!55, label=above:\color{red}{\scriptsize$0$}] (c11) at (-2,1){};
\node[checkSATI, fill=gray!55, label=above:\color{red}{\scriptsize$1$}] (c12) at (2,1){};

\node[var2,  label=above:\scriptsize$0$] (v11) at (-5,0){};
\node[var2,  fill=yellow!35, label=above:\scriptsize$1$] (v12) at (-3,0){};
\node[var2,fill=orange!55, label=above:\scriptsize$2$] (v13) at (-1,0){};
\node[var2,fill=orange!55, label=above:\scriptsize$3$] (v14) at (1,0){};
\node[var2,  fill=yellow!35, label=above:\scriptsize$4$] (v15) at (3,0){};

\node[checkSATI,label=above:\color{red}{\scriptsize$2$}] (c21) at (-4,-1){};
\node[checkPIVOT,label=above:\color{red}{\scriptsize$3$}] (c22) at (0,-1){};

\node[var2, label=above:\scriptsize$5$] (v21) at (-5,-2){};
\node[var,   fill=yellow!35, label=above:\scriptsize$6$] (v22) at (-3,-2){};
\node[var, fill=red!35, label=above:\scriptsize$7$] (v23) at (-1,-2){};
\node[var, fill=red!35,label=above:\scriptsize$8$] (v24) at (1,-2){};
\node[var2,  fill=yellow!35, label=above:\scriptsize$9$] (v25) at (3,-2){};

\node[checkUNSATI,fill=gray!55, label=above:\color{red}{\scriptsize$4$}] (c31) at (-2,-3){};
\node[checkSATI, fill=gray!55, label=above:\color{red}{\scriptsize$5$}] (c32) at (2,-3){};

\node[var2, label=above:\scriptsize$10$] (v31) at (-5,-4){};
\node[var,fill=yellow, label=above:\scriptsize$11$] (v32) at (-3,-4){};
\node[var,fill=red!35,label=above:\scriptsize$12$] (v33) at (-1,-4){};
\node[var,fill=red!35,label=above:\scriptsize$13$] (v34) at (1,-4){};
\node[var2,fill=yellow, label=above:\scriptsize$14$] (v35) at (3,-4){};

\node[checkSATI, label=above:\color{red}{\scriptsize$6$}] (c41) at (-4,-5){};
\node[checkPIVOT, label=above:\color{red}{\scriptsize$7$}] (c42) at (0,-5){};

\node[var2, label=above:\scriptsize$15$] (v41) at (-5,-6){};
\node[var,  fill=yellow!35,label=above:\scriptsize$16$] (v42) at (-3,-6){};
\node[var,fill=orange!55,label=above:\scriptsize$17$] (v43) at (-1,-6){};
\node[var,fill=orange!55,label=above:\scriptsize$18$] (v44) at (1,-6){};
\node[var2,  fill=yellow!35,label=above:\scriptsize$19$] (v45) at (3,-6){};

\node[checkUNSATI, label=above:\color{red}{\scriptsize$8$},fill=gray!55] (c51) at (-2,-7){};
\node[checkSATI, label=above:\color{red}{\scriptsize$9$},fill=gray!55] (c52) at (2,-7){};

\node[var2, label=above:\scriptsize$20$] (v51) at (-5,-8){};
\node[var2,  fill=yellow!35,label=above:\scriptsize$21$] (v52) at (-3,-8){};
\node[var2,  fill=yellow!35,label=above:\scriptsize$22$] (v53) at (-1,-8){};
\node[var2,  fill=yellow!35, label=above:\scriptsize$23$] (v54) at (1,-8){};
\node[var2,  fill=yellow!35, label=above:\scriptsize$24$] (v55) at (3,-8){};

\node[checkSATI, label=above:\color{red}{\scriptsize$10$}] (c61) at (-4,-9){};
\node[checkSATI, label=above:\color{red}{\scriptsize$11$}] (c62) at (0,-9){};

\node[checkDUM, label=above:\color{red}] (ca1) at (4,-1){};
\node[checkDUM, label=above:\color{red}] (ca2) at (4,-5){};
\node[checkDUM, label=above:\color{red}] (ca3) at (-6,-3){};
\node[checkDUM, label=above:\color{red}] (ca4) at (-6,-7){};

\draw (ca1)--(v15) [dashed];
\draw (ca1)--(v25) [dashed];

\draw (ca2)--(v35) [dashed];
\draw (ca2)--(v45) [dashed];

\draw (ca3)--(v21) [dashed];
\draw (ca3)--(v31) [dashed];

\draw (ca4)--(v41) [dashed];
\draw (ca4)--(v51) [dashed];

\draw (c11)--(v12);
\draw (c11)--(v13);
\draw (c12)--(v14);
\draw (c12)--(v15);

\draw (c21)--(v11);
\draw  (c21)--(v12);
\draw  (c21)--(v21);
\draw  (c21)--(v22);

\draw  (c22)--(v13);
\draw  (c22)--(v14);
\draw[  ] (c22)--(v23);
\draw[  ] (c22)--(v24);

\draw[  ] (c31)--(v22);
\draw[  ] (c31)--(v23);
\draw[  ] (c31)--(v32);
\draw[  ] (c31)--(v33);

\draw  (c32)--(v24);
\draw  (c32)--(v25);
\draw  (c32)--(v34);
\draw  (c32)--(v35);

\draw  (c41)--(v31);
\draw  (c41)--(v32);
\draw  (c41)--(v41);
\draw  (c41)--(v42);

\draw[  ] (c42)--(v33);
\draw[  ] (c42)--(v34);
\draw[  ] (c42)--(v43);
\draw[  ] (c42)--(v44);

\draw[  ] (c51)--(v42);
\draw[  ] (c51)--(v43);
\draw  (c51)--(v52);
\draw  (c51)--(v53);

\draw  (c52)--(v44);
\draw  (c52)--(v45);
\draw  (c52)--(v54);
\draw  (c52)--(v55);

\draw  (c61)--(v51);
\draw  (c61)--(v52);

\draw  (c62)--(v53);
\draw  (c62)--(v54);

\end{tikzpicture}   

%% file: tikz_figures/matching_rotated_4.tikz
\begin{tikzpicture}[scale=0.47]
    \tikzstyle{var} = [circle, fill=white, draw=black]
        \tikzstyle{var2} = [circle, fill=white, draw=black]

\tikzstyle{varERR} = [circle, fill=red!55, draw=black]
\tikzstyle{checkUNSATI} = [ fill=white, draw=black,   ]
\tikzstyle{checkSATI} = [ fill=white, draw=black]
\tikzstyle{checkDUM} = [ fill=black, draw=black]

\tikzstyle{checkPIVOT} = [ fill=red, draw=black,   ]

\node[checkSATI,fill=gray!55, label=above:\color{red}{\scriptsize$0$}] (c11) at (-2,1){};
\node[checkSATI, fill=gray!55, label=above:\color{red}{\scriptsize$1$}] (c12) at (2,1){};

\node[var2,  label=above:\scriptsize$0$] (v11) at (-5,0){};
\node[var2,  fill=yellow!35, label=above:\scriptsize$1$] (v12) at (-3,0){};
\node[var2,fill=orange!55, label=above:\scriptsize$2$] (v13) at (-1,0){};
\node[var2,fill=orange!55, label=above:\scriptsize$3$] (v14) at (1,0){};
\node[var2,  fill=yellow!35, label=above:\scriptsize$4$] (v15) at (3,0){};

\node[checkSATI,label=above:\color{red}{\scriptsize$2$}] (c21) at (-4,-1){};
\node[checkPIVOT,label=above:\color{red}{\scriptsize$3$}] (c22) at (0,-1){};

\node[var2, label=above:\scriptsize$5$] (v21) at (-5,-2){};
\node[var,   fill=yellow!35, label=above:\scriptsize$6$] (v22) at (-3,-2){};
\node[var, fill=red!35, label=above:\scriptsize$7$] (v23) at (-1,-2){};
\node[var, fill=blue,label=above:\scriptsize$8$] (v24) at (1,-2){};
\node[var2,  fill=blue!35, label=above:\scriptsize$9$] (v25) at (3,-2){};

\node[checkUNSATI,fill=gray!55, label=above:\color{red}{\scriptsize$4$}] (c31) at (-2,-3){};
\node[checkSATI, fill=blue!35, label=above:\color{red}{\scriptsize$5$}] (c32) at (2,-3){};

\node[var2, label=above:\scriptsize$10$] (v31) at (-5,-4){};
\node[var,fill=yellow, label=above:\scriptsize$11$] (v32) at (-3,-4){};
\node[var,fill=red!35,label=above:\scriptsize$12$] (v33) at (-1,-4){};
\node[var,fill=blue,label=above:\scriptsize$13$] (v34) at (1,-4){};
\node[var2,fill=blue!35, label=above:\scriptsize$14$] (v35) at (3,-4){};

\node[checkSATI, label=above:\color{red}{\scriptsize$6$}] (c41) at (-4,-5){};
\node[checkPIVOT, label=above:\color{red}{\scriptsize$7$}] (c42) at (0,-5){};

\node[var2, label=above:\scriptsize$15$] (v41) at (-5,-6){};
\node[var,  fill=yellow!35,label=above:\scriptsize$16$] (v42) at (-3,-6){};
\node[var,fill=orange!55,label=above:\scriptsize$17$] (v43) at (-1,-6){};
\node[var,fill=orange!55,label=above:\scriptsize$18$] (v44) at (1,-6){};
\node[var2,  fill=yellow!35,label=above:\scriptsize$19$] (v45) at (3,-6){};

\node[checkUNSATI, label=above:\color{red}{\scriptsize$8$},fill=gray!55] (c51) at (-2,-7){};
\node[checkSATI, label=above:\color{red}{\scriptsize$9$},fill=gray!55] (c52) at (2,-7){};

\node[var2, label=above:\scriptsize$20$] (v51) at (-5,-8){};
\node[var2,  fill=yellow!35,label=above:\scriptsize$21$] (v52) at (-3,-8){};
\node[var2,  fill=yellow!35,label=above:\scriptsize$22$] (v53) at (-1,-8){};
\node[var2,  fill=yellow!35, label=above:\scriptsize$23$] (v54) at (1,-8){};
\node[var2,  fill=yellow!35, label=above:\scriptsize$24$] (v55) at (3,-8){};

\node[checkSATI, label=above:\color{red}{\scriptsize$10$}] (c61) at (-4,-9){};
\node[checkSATI, label=above:\color{red}{\scriptsize$11$}] (c62) at (0,-9){};

\node[checkDUM, label=above:\color{red}] (ca1) at (4,-1){};
\node[checkDUM, label=above:\color{red}] (ca2) at (4,-5){};
\node[checkDUM, label=above:\color{red}] (ca3) at (-6,-3){};
\node[checkDUM, label=above:\color{red}] (ca4) at (-6,-7){};

\draw (ca1)--(v15) [dashed];
\draw (ca1)--(v25) [dashed];

\draw (ca2)--(v35) [dashed];
\draw (ca2)--(v45) [dashed];

\draw (ca3)--(v21) [dashed];
\draw (ca3)--(v31) [dashed];

\draw (ca4)--(v41) [dashed];
\draw (ca4)--(v51) [dashed];

\draw (c11)--(v12);
\draw (c11)--(v13);
\draw (c12)--(v14);
\draw (c12)--(v15);

\draw (c21)--(v11);
\draw  (c21)--(v12);
\draw  (c21)--(v21);
\draw  (c21)--(v22);

\draw  (c22)--(v13);
\draw  (c22)--(v14);
\draw[  ] (c22)--(v23);
\draw[  ] (c22)--(v24);

\draw[  ] (c31)--(v22);
\draw[  ] (c31)--(v23);
\draw[  ] (c31)--(v32);
\draw[  ] (c31)--(v33);

\draw  (c32)--(v24);
\draw  (c32)--(v25);
\draw  (c32)--(v34);
\draw  (c32)--(v35);

\draw  (c41)--(v31);
\draw  (c41)--(v32);
\draw  (c41)--(v41);
\draw  (c41)--(v42);

\draw[  ] (c42)--(v33);
\draw[  ] (c42)--(v34);
\draw[  ] (c42)--(v43);
\draw[  ] (c42)--(v44);

\draw[  ] (c51)--(v42);
\draw[  ] (c51)--(v43);
\draw  (c51)--(v52);
\draw  (c51)--(v53);

\draw  (c52)--(v44);
\draw  (c52)--(v45);
\draw  (c52)--(v54);
\draw  (c52)--(v55);

\draw  (c61)--(v51);
\draw  (c61)--(v52);

\draw  (c62)--(v53);
\draw  (c62)--(v54);

\end{tikzpicture}   

%% file: Algorithms/alg_iterative_matching_v2.tex
 \begin{algorithmic}[1]
    \State $\hat{\mathbf{s}} \gets \mathbf{s}$
      \State $\boldsymbol{\gamma}' \gets \boldsymbol{\gamma}$
    \While{$|\hat{\mathbf{s}}|>0$} \label{alg2:stopping}
        \State $i \gets \underset{i \in [1,m]}{\mathrm{arg\ min}}\ \boldsymbol{\gamma}'\ |\ \hat{s}_i=1$ \label{alg2:proxymetric} \Comment{Pivot.}
        \State $\mathbf{t} \gets \hat{\mathbf{s}}$
        \State $t_i \gets 0$
        \State $\boldsymbol{\alpha}(c_i) \gets \mathtt{Shift\_influence}(i)$
        \State $\mathbf{j} \gets j\ |\  t_j>0 \wedge \mathbf{c}_j(c_i)>0$
        \State $j \gets \underset{j \in \mathbf{j}}{\mathrm{arg\ min}}\ \mathbf{c}(c_i) \wedge  \underset{j \in \mathbf{j}}{\mathrm{arg\ min}}\ \boldsymbol{\gamma}'$ \label{alg2:target} \Comment{Target.}
        \State $\boldsymbol{\alpha}(c_j) \gets \mathtt{Shift\_influence}(j)$
        \State $\delta \gets c_j(c_i)$
        \State $\mathbf{f} \gets k\ |\ v_k(c_i)+v_k(c_j)=\delta+1$
        \If{$|\mathbf{f}|=\delta$} \label{alg2:choice}
            \State $\hat{e}_{\mathbf{f}} \gets \hat{e}_{\mathbf{f}} \oplus 1$ \label{alg6:no_degen}
        \Else
            \State Get $\sigma_x$ using (\ref{eq:shift})
            \State Get $\sigma_y$ using (\ref{eq:shift})
            \State $z \gets \phi_c^{-1}(\phi_c(z)+\sigma_x\rho_c)$
            \State $\boldsymbol{\alpha}(c_z) \gets \mathtt{Shift\_influence}(z)$
            \State $\mathbf{f}_1 \gets k\ |\ v_k(c_i)+v_k(c_z)=\Delta x+1$
            \State $\mathbf{f}_2 \gets k\ |\ v_k(c_j)+v_k(c_z)=\Delta y+1$
            \State $\hat{e}_{\mathbf{f}_1,\mathbf{f}_2} \gets \hat{e}_{\mathbf{f}_1,\mathbf{f}_2} \oplus 1$
        \EndIf
        \State $\mathbf{s}' \gets \hat{\mathbf{e}}\cdot \mathbf{H}^{T} \mod{2}$
        \State $\boldsymbol{\gamma}' \gets \mathtt{Shift\_and\_remove}(\cdot,\boldsymbol{\gamma},\mathbf{s}')$
        \State $\hat{\mathbf{s}} \gets \mathbf{s} \oplus \mathbf{s}'$ \label{alg2:residual}
        \EndWhile
    \State \Return $\hat{\mathbf{e}}$
  \end{algorithmic}